\documentclass [a4paper, 11 pt] {article}

\usepackage{amsmath}
\usepackage{amsfonts}
\usepackage{amsthm}
\usepackage{amscd}

\renewcommand\Re{\hbox{\rm Re}\,}
\renewcommand\Im{\hbox{\rm Im}\,}
\newcommand\Rn{{\mathbb R}}
\newcommand\Cn{{\mathbb C}}
\newcommand\ket[1]{|#1\rangle}
\newcommand\bra[1]{\langle #1 |}
\newcommand\braket [2] {\langle #1 | #2 \rangle}
\newcommand\me [3]{\langle #1 | #2 | #3 \rangle}
\newcommand\kpsi[1]{\ket{\psi(#1)}}
\newcommand\kn[1]{\ket{n(#1)}}
\newcommand\R{{\bf R}}
\newcommand\A{{\bf A}}\newcommand\D{{\bf D}}\newcommand\F{{\bf F}}
\newcommand\Id{{\rm I}}
\newcommand\Am{{\rm A}}
\newcommand\Amb{ \textsf{\textbf{ A}}} 

\newcommand\V{{\bf V}}

\newcommand\Vmb{ \textsf{\textbf{ V}}}
\newcommand\Hm{{\rm H}}
\newcommand\B{{\bf B}}
\newcommand\E{{\bf E}}
\renewcommand\L{{\bf L}}
\renewcommand\l{{\bf l}}
\newcommand\dett{{\rm det}}
\renewcommand\P{{\bf P}}
\renewcommand\S{{\bf S}}
\newcommand\W{{\bf W}}
\newcommand\M{{\bf M}}
\newcommand\rot{{\rm rot}}
\newcommand\bnabla{{\boldsymbol \nabla}}
\newcommand\gradR{\bnabla}
\newcommand\gradk{\bnabla_\k}
\newcommand\defn{\overset{\text{defn}}{=}}
\renewcommand\r{{\bf r}}
\newcommand\s{{\bf s}}
\newcommand\p{{\bf p}}
\renewcommand\k{{\bf k}}
\renewcommand\v{{\bf v}}
\newcommand\vpsi{{\boldsymbol \psi}}
\renewcommand\Id{{\rm I}}
\newcommand\sgn{\hbox{\rm sgn}\,}

\newcommand\ph{{\bar{\cal H}}}
\newcommand\ehat{{\bf \hat e}}
\newcommand\nhat{{\bf \hat n}}
\newcommand\bhat{{\bf \hat b}}
\newcommand\dhat{{\bf \hat d}}
\newcommand\that{{\bf \hat t}}

\newcommand\xhat{{\bf \hat x}}
\newcommand\yhat{{\bf \hat y}}
\newcommand\zhat{{\bf \hat z}}
\newcommand\pauli{{\bsigma}}
\newcommand\ma{m_{\alpha}}
\newcommand\ra{\r_{\alpha}}
\newcommand\va{\v_{\alpha}}
\newcommand\Va{\V_{\alpha}}
\newcommand\Ra{\R_{\alpha}}
\newcommand\calR{{\cal R}}
\newcommand\tkn{TKN$^2$}
\newcommand\Tr{\,\text{Tr}\,}
\newcommand\bsigma{{\boldsymbol \sigma}}
\newcommand\bomega{{\boldsymbol \omega}}
\newcommand\bOmega{{\boldsymbol \Omega}}
\newcommand\bphi{{\boldsymbol \phi}}

\numberwithin{equation}{section}

\begin{document}

\title{TOPOLOGICAL PHASE EFFECTS}[Topological phase effects]

\author{JM Robbins} [JM Robbins]

\thanks{ Basic Research Institute in the Mathematical Sciences\\
Hewlett-Packard Laboratories Bristol\\
Filton Road, Stoke Gifford, Bristol BS12 6QZ\\
and\\
School of Mathematics\\
University Walk, Bristol BS8 1TW}
\maketitle

\begin{abstract}
Quantum eigenstates undergoing cyclic changes acquire a phase factor of
geometric origin.  This phase, known as the Berry phase, or the
geometric phase, has found applications in a wide range of
disciplines throughout physics, including atomic and molecular
physics, condensed matter physics, optics, and classical dynamics.
In this article, the basic theory of the geometric phase is presented along
with a number of representative applications.

The article begins with an account of the geometric phase for cyclic
adiabatic evolutions.  An elementary derivation is given along with a
worked example for two-state systems. The implications of
time-reversal are explained, as is  the fundamental connection
between the geometric phase and energy level degeneracies.  We also
discuss methods of experimental observation.  A brief account is given
of geometric magnetism; this is a Lorenz-like force of geometric
origin which appears in the dynamics of slow systems coupled to fast
ones.

A number of theoretical developments of the geometric phase are
presented.  These include an informal discussion of fibre bundles, and
generalizations of the geometric phase to degenerate eigenstates (the
nonabelian case) and to nonadiabatic evolution.  There follows an
account of applications.  Manifestations in classical physics include
the Hannay angle and kinematic geometric phases.  Applications in
optics concern polarization dynamics, including the theory and
observation of Pancharatnam's phase.  Applications in molecular
physics include the molecular Aharonov-Bohm effect and nuclear
magnetic resonance studies.  In condensed matter physics, we discuss
the role of the geometric phase in the theory of the quantum Hall
effect.

\end{abstract}
\vfill\eject

{
\obeylines
\ \phantom{\ref{basic}}\ {\bf Introduction}

{\ref{basic}\ {\bf Basic Account of the Geometric Phase}\hfill3}\\
{\ref{bas.adiabatic}\ {\it Quantum adiabatic theorem}\hfill4}\\
{\ref{bas.geophase}\ {\it The geometric phase}\hfill4}\\
{\ref{bas.spin}\ {\it Two-state systems}\hfill5}\\
{\ref{bas.degen}\ {\it Degeneracies}\hfill7}\\
{\ref{bas.tr}\  {\it Time-reversal symmetry}\hfill9}\\
{\ref{bas.exper}\ {\it Experimental observation}\hfill10}\\
{\ref{bas.geomag}\ {\it Geometric magnetism}\hfill11}\\

{\ref{theory}\ {\bf Theoretical Developments}\hfill13}\\
{\ref{theory.hol}\ {\it Geometric phase as holonomy}\hfill13}\\
{\ref{theory.hol.fb}\ {\it Fibre bundles.}\hfill13}\\
{\ref{theory.hol.con}\ {\it Connections and curvature.}\hfill14}\\
{\ref{theory.hol.geo}\ {\it Geometric phase revisited.}\hfill15}\\
{\ref{theory.hol.chern}\  {\it Chern classes.}\hfill15}\\
{\ref{theory.nonab}\ {\it Degenerate eigenstates and nonabelian geometric phases}\hfill16}\\
{\ref{theory.nonad}\ {\it Nonadiabatic evolution}\hfill19}\\
{\ref{theory.metric}\ {\it Geometric metric tensor}\hfill21}\\
{\ref{theory.beyad}\ {\it Beyond adiabaticity}\hfill22}\\
{\ref{corr}\ {\it Corrections to the geometric phase}\hfill22}\\
{\ref{amplitude}\ {\it The geometric amplitude}\hfill24}\\

{\ref{classical}\ {\bf Classical Systems}\hfill26}\\
{\ref{hannay}\ {\it The Hannay angle}\hfill26}\\
{\ref{hannay.cat}\ {\it The classical adiabatic theorem.}\hfill26}\\
{\ref{hannay.bas}\ {\it Basic account of the Hannay angle.}\hfill27}\\
{\ref{hannay.ex}\ {\it Examples.}\hfill28}\\
{\ref{hannay.sc}\ {\it Semiclassical limit.}\hfill29}\\
{\ref{hannay.react}\ {\it Classical reaction forces.}\hfill30}\\
{\ref{classical.kin}\ {\it Symmetry and kinematic geometric phases}\hfill31}\\
{\ref{kin.cats}\ {\it Falling cats and swimming paramecia.}\hfill31}\\
{\ref{kin.red}\ {\it Geometric phases and symmetry reduction.}\hfill33}\\
{\ref{Opt}\ {\it Optimization and control.}\hfill35}\\

{\ref{optics}\ {\bf Optics}\hfill35}\\
{\ref{optics.coiled}\ {\it Geometric phase of coiled light.}\hfill36}\\
{\ref{optics.panch}\ {\it Pancharatnam's phase.}\hfill37}\\

{\ref{applications}\ {\bf Molecular and Condensed Matter Physics}\hfill40}\\
{\ref{pseudo}\ {\it Pseudorotation in triatomic molecules}\hfill40}\\
{\ref{nmr}\ {\it Nuclear magnetic resonance}\hfill42}\\
{\ref{qhe}\ {\it The quantum Hall effect}\hfill44}\\
{\ \phantom{\ref{basic}}\ {\bf Glossary}\hfill48}\\
{\ \phantom{\ref{basic}}\ {\bf Works Cited}\hfill48}\\
{\ \phantom{\ref{basic}}\ {\bf Further Reading}\hfill50}\\
}
\vfill\eject
\section{Introduction}\label{sec: intro}
\noindent In a paper published in 1984, Sir Michael Berry (1984) found that a
quantum system undergoing adiabatic evolution acquires a phase factor
of purely geometrical origin.  This discovery, now called the Berry
phase, or the geometric phase, initiated a tremendous surge of
research in a range of disciplines across physics.  Berry and
subsequent workers quickly established connections to a number of
apparently disparate phenomena, including amongst others the
Aharonov-Bohm effect (Aharonov and Bohm 1959), its molecular physics
analogue (Mead and Truhlar 1979), polarization optics (Pancharatnam
1956), the quantum Hall effect (Thouless et al.\ 1982), and the Foucault
pendulum (Eco 1989).  Beyond providing a unified description of
these phenomena, the geometric phase led to predictions of new
effects, subsequently observed experimentally, and a number of
substantial extensions and generalizations.  Among these, the
discovery by JH Hannay (1985) of an analogous effect in classical
mechanics, the Hannay angle, stimulated much new research in purely
classical physics.

The mathematical phenomenon which underlies the geometric phase is
holonomy.  Holonomy describes transformations in a quantity induced by
cyclic changes in the variables which control it.  An example from
geometry is the rotation of vectors parallel transported along closed
paths on a curved surface.  The mathematics of holonomy and curvature
has long played a role in physics through general relativity and
quantum field theory.  With Berry's work came the appreciation of a
broader scope of application.  Whenever a physical system is divided
into parts and one attempts to describe a subsystem in isolation, the
influence of the others is manifested through geometric phase effects.
The pervasiveness througout physics of this reductive procedure
accounts for the ubiquity of the geometric phase.  Introducing a
collection of major papers in the field, A Shapere and F Wilczek(1989)
wrote, ``We believe that the concept of a geometric phase, repeating
the history of the group concept, will eventually find so may
realizations and applications in physics that it will repay study for
its own sake, and become part of the {\it lingua franca}.''

An introductory account of the geometric phase is given in
Section~\ref{basic}.  This is followed in Section~\ref{theory} by a
discussion of several theoretical developements.  Applications to
classical systems, optics, and molecular and condensed matter physics
are given in Sections~\ref{classical}--\ref{applications}.  The  topics
chosen for discussion meant to be representative rather than
comprehensive.  Among several omissions, we mention the theory of
fractional statistics, which is covered extensively in the volume by
Shapere and Wilczek (1989a), and applications to elementary particle
physics, which are reviewed by Aitchison (1987).  The historical
development of the subject is discussed  by Berry (1990a). \vfill\eject

\section{Basic Account of the Geometric Phase\label{basic}}
The geometric phase was discovered by Sir Michael Berry in 1983 in the
course of a critical reexamination of the adiabatic theorem in quantum
mechanics.  This theorem says, in essence, that if a quantum system is
changed sufficiently slowly, it does not undergo transitions.  Thus a
system initially in an eigenstate of a slowly changing Hamiltonian
remains in the evolved eigenstate.  Berry considered the overall phase
accumulated by the changing eigenstate and found an unexpected
contribution.  This extra phase is of a geometrical nature, in that it
does not depend on the rate at which the adiabatic process is carried
out.

One way that a slowly varying Hamiltonian can arise physically is when
the system under consideration interacts with another, much slower
system.  This leads to the consideration of the quite general problem
of two coupled systems evolving on different time scales.  In this
context, the geometric phase describes an effect on the fast system
produced by the slow one.  As action in physics is accompanied by
reaction, there is a reciprocal effect on the slow system generated by
the fast.  This takes the form of an effective vector potential in the
slow Hamiltonian; the associated Lorentz-like force is called
geometric magnetism.  The same geometry underlies both
geometric magnetism and the geometric phase.

This geometrical structure is intimately connected to degeneracies
in the energy spectrum.  A typical Hamiltonian has a nondegenerate
spectrum (excepting systematic degeneracies due to symmetry, which
are discussed in Section \ref{theory.nonab}).  Varying a single
parameter (eg time itself in the case of a time-dependent
Hamiltonian) is unlikely to produce any degeneracies; typically
two or three parameters (the number depends on whether
time-reversal symmetry is present or not) must be adjusted
independently in order to find one. Nevertheless, the presence of
degeneracies in Hamiltonians which are, in some sense, near the
Hamiltonians of the adiabatic process, are largely and sometimes
wholly responsible for the geometric phase. This connection
underlies several applications, including to the quantum Hall
effect (Section~\ref{qhe}).

\subsection{Quantum adiabatic theorem\label{bas.adiabatic}}
Adiabatic invariants are quantities which are nearly conserved when
the environment of an isolated system is slowly changed (slowly means
in comparison with the system's internal dynamics).  Adiabatic
theorems provide estimates as to how accurately this conservation is
respected.
In quantum mechanics, the adiabatic invariants are the probabilities
$P_n = |\braket n \psi |^2$ to be in energy eigenstates.  These are
constant for stationary Hamiltonians, but vary if $H$ is
time-dependent.  The quantum adiabatic theorem states that as $dH/dt$
goes to zero, so do the probabilities of transitions between
eigenstates (the rate depends on how smooth $dH/dt$ is), so that the
$P_n$ are nearly constant.  Therefore, within the adiabatic
approximation, dynamics under a slowly varying Hamiltonian $H(t)$ is
essentially no more complicated than under a stationary one.  In
particular, if $\ket{\psi}$ begins in an eigenstate, it remains in one
( nonadiabatic transitions are discussed in
Section~\ref{theory.beyad}), and its evolution is determined up to a
phase factor.

We shall now examine this phase.  Suppose that at $t=0$, $\ket{\psi}$
is in the $n$th eigenstate $\ket{n}$ (assumed nondegenerate) of $H$.
For its subsequent evolution we may write
  \begin{equation}\kpsi{t} \approx
\exp\left(-i/\hbar\int^t E_n(\tau)d\tau\right)
\kn{t}.\label{ansatz}\end{equation} In so doing we have explicitly
identified part of the overall phase, the time integral of the
frequency $E_n/\hbar$, which is called the {\it dynamical phase}.
It is just the phase one would expect for a solution of the
Schr\"odinger equation.  Any remaining phase is contained
implicitly in the eigenstate $\kn{t}$.  Its phase is determined by
requiring that $\me n {(i\hbar \partial_t - H)} {\psi}$ vanish (it
would automatically if $\kpsi{t}$ were an exact solution of the
Schr\"odinger equation).  This leads to the condition
\begin{equation}\braket {n} {\dot n} = 0,\label{parallel}\end{equation} which determines
the phase of $\kn{t}$ given an initial choice at $t = 0$.

\subsection{The geometric phase\label{bas.geophase}} What Berry
discovered is that condition (\ref{parallel}) yields a path-dependent
phase factor.  This phase factor is particularly interesting when the
Hamiltonian is cycled, ie made to return to itself after some time.
Then it depends only on the cycle itself, and not, in contrast to the
dynamical phase, on the rate at which the Hamiltonian is cycled.  For
this reason it is called, as well as the {\it Berry phase}, the {\it
geometric phase}.

To discuss this further, it helps to formulate the problem
slightly differently.  Suppose the time dependence of the
Hamiltonian is expressed through its dependence on some external
parameters $\R = (X,Y,Z)$, which are made to vary slowly in time
along a path $\R(t)$. Then the eigenstates $\kn{\R}$ depend on
$\R$ as well, and (assuming no degeneracies) are determined for
each $\R$ up to an overall phase. Let us choose this phase to vary
smoothly but otherwise arbitrarily with $\R$.  By fixing the phase
independently of the dynamics, we may no longer satisfy the
condition (\ref{parallel}) along every path $\R(t)$. We therefore
allow for an additional phase factor $\exp (i\gamma_n(t))$ in the
adiabatic ansatz (\ref{ansatz}), which becomes
  \begin{equation}\kpsi{t} \approx \exp\left(-i/\hbar\int_0^t
E_n(\R(\tau))\,d\tau\right) \exp (i\gamma_n(t))
\kn{\R(t)}.\label{ansatz2}\end{equation} The condition $\me n
{(i\hbar
\partial_t - H)} {\psi} = 0$ now yields an evolution equation for
$\gamma_n$. This may be written in the form $\dot \gamma_n(t) =
-\A_n(\R(t))\cdot\dot \R(t)$, where $\A_n(\R)$ is a vector field
in parameter space given by \begin{equation}\A_n(\R) = \Im \braket
{n(\R)} {\gradR n(\R)},\label{vecpot}\end{equation} or
equivalently, in terms of eigenfunctions $\psi_n(x,\R) = \braket
{x} {n(\R)}$, by $\A_n(\R) = \Im \int \psi^*_n(x,\R)\gradR
\psi(x,\R)dx$ (here $\gradR$ acts on the parameters $\R$).

If the Hamiltonian is cycled around a closed curve $C$ in
parameter space, then $\kpsi{t}$ acquires a geometric phase
\begin{equation}\gamma_n(C) = -\int_0^T \A_n\cdot\dot \R\, dt = \oint_C
\A_n\cdot d\R. \label{geo}\end{equation} Its expression as a line
integral reveals its geometrical character; the geometric phase
depends only on the cycle $C$, and not on the manner in which the
cycle is traversed in time. Using Stokes' theorem, $\gamma_n$ can
be expressed as \begin{equation}\gamma_n(C) = -\int_S
\V_n(\R)\cdot d\S,\label{geo2}\end{equation} the flux of the
vector field $\V_n = \gradR\times \A_n$ through a surface $S$ in
parameter space whose boundary is $C$.  $V_n$ is given by
\begin{equation}\V_n = \Im \bra{\gradR n}\times \ket{\gradR
n}.\label{magfld}\end{equation} From (\ref{geo2}) it is evident
that only cycles which enclose nonzero area can have nontrivial
geometric phases. For example, $\gamma_n(C)$ vanishes for cycles
consisting of a path and its retrace.

It is easy to see that the geometric phase does not depend on the
choice of phase for $\ket{n(\R)}$.  Indeed, under the change of phase
  \begin{equation}\kn{\R} \rightarrow \exp
  (i\chi(\R))\kn{\R},\label{gt}\end{equation}
$\A_n(\R)$ transforms according to $\A_n(\R) \rightarrow \A_n(\R)
+ \gradR \chi(\R)$; this  shift by a perfect gradient  leaves its
integral round a closed loop, $\gamma_n$, and its curl, $\V_n$,
unchanged. Thus, the geometric phase is an intrinsic property of
the family of eigenstates $\ket{n(\R)}$.

It is useful to draw an analogy with electromagnetism, in which $\A_n$
corresponds to the vector potential and $\V_n$ to the magnetic field.
The geometric phase $\gamma_n$ corresponds to a magnetic flux.  The
change of phase
(\ref{gt}) corresponds to a gauge transformation, which shifts the
vector potential by a gradient and leaves the magnetic field
unchanged.

Substituting the first-order perturbation expansion
\begin{equation}\ket{\gradR n} = \sum_{m\ne n} \frac{\me {m}{\gradR H} {n}} {
(E_n-E_m)}\ket{m} + \braket {n} {\gradR n}\,
\ket{n}\label{pert}\end{equation} into \eqref{magfld} leads to a
useful formula, \begin{equation}\V_n = \Im \sum_{m\ne n} \frac{\me
{n} {\gradR H} {m} \times \me {m} {\gradR H}
{n}}{(E_n-E_m)^2}.\label{alt}\end{equation} In this form, the
independence of $\V_n$ on the eigenstate phases is manifest.
Another formula with this property is \begin{equation}\V_n =
\Tr\,(
\rho_n\gradR\rho_n\times\gradR\rho_n),\label{Avron}\end{equation}
where $\rho_n = \ket{n}\bra{n}$ is the density operator.  It too
is useful in applications.

One of the applications discussed by Berry (1984) is to the
Aharonov-Bohm effect (Aharonov and Bohm 1959).  The Aharonov-Bohm
effect describes the dependence of quantum mechanics on
electromagnetic fields in regions of space which are physically
excluded.  This dependence is mediated through the vector and scalar
potential, whose relation to the electric and magnetic fields is
nonlocal.  The most celebrated example concerns an electron moving in
the vicinity of an inpenetrable perfect solenoid of infinite extent.
Since outside the solenoid the magnetic field is zero, the classical
mechanics of the electron is insensitive to any magnetic flux inside
the solenoid.  But the quantum mechanical behaviour of the electron
does depend on the interior flux, as Y Aharonov and D Bohm showed by
calculating its scattering cross-section (a prediction subsequently
experimentally confirmed by Chambers (1960)).

The Aharonov-Bohm effect can be understood in terms of
interference between de Broglie waves passing on opposite sides of
the solenoid. The phase difference between the two waves includes
the integral \begin{equation}\gamma_{AB} = \frac{e}{\hbar c}
\oint_C \A^{(s)}(\R)\cdot d\R\label{abphase}\end{equation} of the
vector potential $\A^{(s)}(\R)$ of the solenoid round a loop $C$
enclosing it.  (In \eqref{abphase}, $c$ is the speed of light and
$-e$ the electron charge.)  The Aharonov-Bohm phase $\gamma_{AB}$
is just $2\pi$ times the magnetic flux in the solenoid measured in
units of the flux quantum $hc/e$.

It may be regarded as a geometric phase in the following manner.  We
imagine the electron confined to a large box, centred at $\R$, which
is not penetrated by the solenoid.  The eigenstates of the electron
$\ket {n(\R)}$ depend parametrically on $\R$; changes in $\R$ induce
gauge transformations in the eigenfunctions.  One finds that $\Im
\braket {n(\R)} {\gradR n(\R)} = e/\hbar c \A^{(s)}$, so that in this
case the geometric vector potential \eqref{vecpot} coincides with the
physical vector potential of the solenoid.  On being transported (not
necessarily adiabatically in this case) round the solenoid, the
electron eigenstates acquire a geometric phase which is precisely the
Aharonov-Bohm phase.

\subsection{Two-state systems\label{bas.spin}}
Two-state systems provide a canonical example of the geometric
phase. The requisite calculations are easily carried out, and the
main features of the general case are represented.  The
parameterized family of Hamiltonians can be written in the form
\begin{equation}H(\R) = E(\R)\Id + \F(\R)\cdot\bsigma,\label{H2}\end{equation} where
$\Id$ is the identity and $\bsigma$ are the Pauli spin matrices.
The term $E(\R)\Id$ shifts the energy by a constant and affects
neither the eigenstates nor their geometric phases.  Both of these
are determined by the field $\F(\R)$.

An illustrative example is a spin-1/2 particle in a uniform
magnetic field.  The parameters $\R$ are then just the components
of the field, so that $\F(\R) = \frac12 \mu \R$, where $\mu$ is
the magnetic moment.  The eigenstates \begin{equation}\ket{+(\R)}
= \left(
  \begin{array}{c}
    \cos\theta/2 \\
    e^{i\phi}\sin\theta/2 \\
  \end{array}
\right), \quad
     \ket{-(\R)} =
     \left(
       \begin{array}{c}
         -\sin\theta/2\\
         e^{i\phi}\cos\theta/2  \\
       \end{array}
     \right)
\label{spins}\end{equation} depend on the polar angles
$(\theta,\phi)$ of $\R$, and the energy levels $E_{\pm}(\R) =
\pm\frac12 \mu \hbar R$ on its magnitude. Straightforward
calculation using \eqref{vecpot} and \eqref{magfld} yields
\begin{equation}\A_{\pm}(\R) = \pm \frac12 \frac{1-\cos\theta}{
R\sin\theta}{\bf\hat\bphi},\quad \V_{\pm}(\R) = \pm \frac12
\frac{\R}{ R^3}.\label{spinflds}\end{equation} Thus, $\A_{\pm}$
and $\V_{\pm}$ are respectively the vector potential and magnetic
field of a magnetic monopole of strength $\pm 2\pi$ at the origin.
Under the influence of a slowly varying magnetic field $\R(t)$
taken through a cycle $C$, a spin initially in an up-state returns
to itself with a dynamical phase $-\mu/2\hbar\, \int R d\tau$ and
a geometric phase $\gamma_+(C) = -\int_S \V_+ \cdot d\S$.  Gauss's
law implies that \begin{equation}\gamma_+(C) = -\frac12\,\times
\hbox{\rm solid angle subtended by C}.
\label{solidangle}\end{equation} This spin-1/2 geometric phase has
been observed in neutron spin rotation experiments by Bitter and
Dubbers (1987). A special case is when $C$ is confined to a plane
containing the origin.  Then its solid angle is $2\pi$ times the
number of times $m$ the origin is enclosed, and the corresponding
phase factor is $(-1)^m$.  This corresponds to the well-known sign
change of spin-1/2 particles under rotations through $2\pi$.

It is straightforward to generalize these results to the case of
arbitrary $\F(\R)$ in \eqref{H2}.  The flux field is then given by
\begin{equation}\V_{\pm}(\R) = \pm \frac12 \epsilon_{ijk} \frac{F_i \gradR
F_j \times \gradR F_k}{F^3}\label{spin2}\end{equation}
($\epsilon_{ijk}$ is the antisymmetric Levi-Cevita symbol), and it
has monopole sources at points where $\F$ vanishes.  The monopoles
have strength $\pm 2\pi\sigma$, where $\sigma$ is the sign of the
(assumed nonvanishing) determinant $\det \partial F_i/\partial
R_j$ evaluated where $\F = 0$.

As is familiar from Dirac's analysis of magnetic monopoles (Dirac
1931), there is necessarily a string of singularities of the
vector potential attached to the monopole.  The position of the
string depends on the choice of gauge, and Dirac's quantization
condition for electric charge follows from demanding that these
choices be physically indistinguishable.  In the context of the
two-state geometric phase, the monopole strings of $\A_\pm(\R)$
lie along discontinuities in the phases of $\ket{\pm(\R)}$.  When
the eigenstates are chosen as in \eqref{spins}, these
discontinuities lie along the negative $z$ axis.  By varying the
phase of the eigenstates (as in \eqref{gt}), the string can be
made to pass in and out of a given cycle $C$.  As it does, the
geometric phase $\gamma_\pm(C)$ changes by $\pm 2\pi$.  Therefore,
the geometric phase factor $\exp (i\gamma_\pm(C))$ remains
unchanged, and there are no observable consequences.

\subsection{Degeneracies\label{bas.degen}}
In the spin-1/2 example, the flux fields $\V_\pm(\R)$ are generated by
point sources of strength $\pm2\pi$ at the origin $\R =0$.  What makes
the origin special is that it is the only point in parameter space
where the energy levels are degenerate. This relationship between the
geometric phase and energy level degeneracies turns out to be quite
general.  In the absence of symmetries, a typical Hamiltonian in a
parameterized family has a nondegenerate spectrum.  However, at
certain points in parameter space, belonging to the so-called {\it
degenerate set}, two (or more) energy levels become degenerate.  In
the absence of time-reversal symmetry (the time-reversal symmetric
case is discussed in the following section), the dimension of the
degenerate set is typically three less than the dimension of parameter
space itself. Thus, for the three-parameter families $H(\R)$ we have
been considering, degeneracies occur at isolated points.  We have
already seen this to be the case in the spin-1/2 example.  (Note that
there, the magnetic field is responsible for breaking time-reversal
symmetry.)

More generally, let us consider the behaviour of the flux field
$V_n(\R)$ in the neighbourhood of a degeneracy $\R_*$ between two
eigenstates, say the $n$th and $n-1$st, of an arbitrary family of
Hamiltonians. As $E_n(\R) - E_{n-1}(\R)$ is small near $\R_*$,
perturbation theory (cf \eqref{pert})  implies that the dominant
component of $\ket{\gradR n}$  lies along $\ket{n-1}$. Therefore,
for the purposes of approximating $\V_n$  (which is defined in
terms of $\ket{\gradR n}$), the Hamiltonian can be restricted to
the nearly-degenerate subspace spanned by the states
$\ket{1}\defn\ket{n(\R_*)}$ and $\ket{2}\defn\ket{n-1(\R_*)}$. One
obtains in this way a two-state system of the form \eqref{H2},
where the field $\F(\R)$ is given by \begin{equation} F_x = \Re
\me {1} {H(\R)} {2},\  F_y = \Im \me {1} {H(\R)} {2},\ F_z =
\frac{\me {1} {H(\R)} {1} - \me {2} {H(\R)}
{2}}{2}.\label{Ffld}\end{equation}

In the neighbourhood of $\R_*$, $\V_n(\R)$ is given approximately
by \eqref{spin2}.  There it has a single monopole source at the
degeneracy $\R_*$ of strength $\pm 2\pi$.  We may apply a similar
analysis in the neighbourhood of each degeneracy of $\ket{n}$ to
obtain \begin{equation}\gradR\cdot \V_n = \sum_{\R_*} \pm 2\pi
\delta(\R - \R_*). \label{degen}\end{equation} Thus, the flux
field for the $n$th eigenstate has monopole sources of strength
$\pm 2\pi$ at points where $\ket{n}$ is degenerate.  (Their sign
depends on whether the degeneracy is with the state above or
below, and also on the sign of the determinant $\det \partial
F_i(\R_*)/\partial R_j$.) $\V_n$ may in addition have a purely
divergenceless contribution.

It follows that the flux of $\V_n$ through a closed surface in
parameter space is $2\pi$ times the number of degeneracies, counted
with their signs, contained within.

\subsection{Time-reversal symmetry\label{bas.tr}}
The presence of time-reversal symmetry imposes constraints on the
geometric phase.  These  are strongest when the eigenstates
themselves are time-reversal symmetric, so we
consider such cases first.
Then the wave functions $\psi_n(x,\R) = \braket x {n(\R)}$ can be
chosen to be real, and the geometric phase factor is necessarily a
sign $\pm1$, so that $\gamma_n$ is an integer multiple of $\pi$.
In this case, the geometric phase depends only on the topology
of the closed curve, not on its geometry.  As $C$ is deformed,
unless $\gamma_n(C)$ changes discontinuously, it
necessarily remains constant.
It can change (discontinously) when $C$ passes through
degeneracies (points $\R_*$ where $\ket n$ is degenerate).  For systems
with time-reversal symmetry, the dimension of this degenerate set is
typically two less than the dimension of parameter space,
so that it consists of lines in three-dimensional
parameter space, or points in two-dimensional parameter space.  The
geometric phase $\gamma_n(C)$ is $\pi$ times the number of
degeneracies enclosed by $C$, counted with the appropriate sign.
An important consequence is that the flux field $\V_n$ has delta-function
singularities at the energy level degeneracies and is zero elsewhere.

The spin-1/2 example of Section~\eqref{bas.spin} provides an
illustration if we fix $Y$ (the y-component of the magnetic field) to
be zero.  Then the spin Hamiltonian $\frac12\mu \B.\bsigma$ is real
symmetric rather than complex Hermitian, and we can take its
eigenvectors to be real.  The parameter space is the $XZ$ plane, and
the solid angles subtended by curves in this plane are multiples of
$2\pi$. The geometric phase $\gamma_{\pm}(C)$ is just $\pm\pi$
times the number of times $C$ encircles the origin.
Another example concerns the electronic ground state of triatomic molecules
$X_3$ (cf Section~\ref{pseudo}).

In cases where the eigenstates are themselves not time-reversal symmetric,
and instead occur in pairs related by time-reversal, the constraints
on the geometric phases are much weaker.  They are no longer
necessarily multiples of $\pi$ (ie, the phase factors are no longer
necessarily real). However, geometric phases for states related by
time-reversal necessarily have opposite signs.

\subsection{Experimental observation\label{bas.exper}}
Since the physical properties of a quantum system does not depend on its
overall phase, one could wonder whether the geometric phase is observable.
Numerous experiments have shown that it is.

Most direct experimental
studies of the geometric phase belong to one of three types.  The
first is the `one state, two Hamiltonian' experiment.
An ensemble of states initially in an energy eigentate is
coherently divided into two streams.
The first evolves under a constant Hamiltonian, while
the Hamiltonian driving the second is cyclically varied.
The two streams are then recombined and allowed to interfere, and their
relative phase  determined.  The relative phase
includes both dynamical and geometric contributions, and to extract
the geometric term
the dynamical phase must either be made to vanish,
or else  subtracted off explicitly.
Examples are the optical polarization experiments discussed
in Section~\ref{optics.panch}.

The second type is the `two state, one Hamiltonian' experiment.  Here an
initial state $\ket{\psi_i}$ is prepared in a superposition $\alpha_i
\ket {m} + \beta_i \ket {n}$ of\ energy eigenstates. The Hamiltonian
governing its evolution is taken through an adiabatic cycle.  The
final state $\ket{\psi_f} = \alpha_f \ket {m} + \beta_f \ket {n}$ is
physically distinguishable from the initial one, as the relative phase
of the superposed eigenstates has changed.  Part of this (observable)
change in the relative phase involves the difference $\gamma_m -
\gamma_n$ in geometric phases.
The
observation of optical activity in coiled fibres by Tomita and Chiao
(1986) (see Section~\ref{optics.coiled}) provides an example of this
second type.

In the third type of experiment, the adiabatic cycle is performed not
once but repeatedly, with period $T$.  With each cycle, $\ket{n}$ acquires
a phase $\phi = -\int E_n(\R(t)dt/\hbar + \gamma_n$, so that there appears in
its mean angular frequency $
\omega = \phi/T$ a geometric shift $\delta \omega_g = \gamma_n/T$.  This
frequency shift can be detected in spectroscopic measurements, and is
the basis for nuclear magnetic resonance studies of the geometric
phase (cf Section~\ref{nmr}).

\subsection{Geometric magnetism\label{bas.geomag}}
So far we have been regarding the evolution of the parameters $\R$ as
externally prescribed.  Now let us suppose the parameters of the
Hamiltonian $H(\R)$ are dynamical variables in their own right.  For
example, they could be the coordinates of a heavier (and therefore
slower) system to which the original one, the fast system, is coupled.
For definiteness, we take the Hamiltonian of the coupled system to be
of the form $P^2/2M + H(\R)$, where $\P$ is the momentum of the slow
system, and the dependence on the fast coordinates and momenta is left
implicit in $H(\R)$. Both systems are to be treated quantum
mechanically.  The geometric phase describes an effect on the fast
system due to the slow.  There is a corresponding influence
on the slow system due to the fast.

This is the natural setting for the Born-Oppenheimer approximation, in
which the fast and slow systems are the electrons and nuclei
of a molecule.  (For consistency with the preceding discussion,
we denote the nuclear coordinates by a single vector $\R$; it is easy
to adapt the notation to accomodate more than one nucleus.)
By assuming that the electronic state rapidly adjusts
to changes in the nuclear configuration, one is led to look for
approximate eigenfunctions of the form $\Psi_n(\r,\R) =
\psi_n(\R)\phi_n(\r,\R)$, where the electronic wavefunction $\phi_n(\r,\R)$
is an eigenstate with energy $E_n(\R)$ of the
electronic Hamiltonian $H(\R)$, in which the
nuclear coordinates appear as parameters.
The nuclear wavefunction
$\psi_n(\R)$ is an eigenstate of the nuclear Hamiltonian $-\hbar^2
{\gradR}^2/2M + E_n(\R)$; the electronic energy plays the role of a
potential.  The Born-Oppenheimer
approximation consists of neglecting the
weak variation of the electronic wave function
$\phi_n(\r,\R)$ with the nuclear coordinates.
(It is essentially the adiabatic approximation \eqref{ansatz} adapted to
coupled systems.)

This was the context for seminal work in this subject by CA Mead
and D Truhlar in 1979. The primary objective of this work, which
anticipated certain aspects of the geometric phase, was to derive
corrections to the Born-Oppenheimer approximation.  Such
corrections appear when the momentum operator $\P = -i\hbar
\gradR$ is applied to the putative eigenstate $\psi_n(\R)\kn{\R}$.
The result, $ -i \hbar\left(\gradR \psi_n\ket n + \psi_n \ket
{\gradR n}\right)$, is no longer proportional to the electronic
eigenstate $\kn{\R}$.  However, if we neglect electronic
components orthogonal to $\ket n$, we obtain $-i\hbar\left(\gradR
\psi_n\ket n + \psi_n\braket {n} {\gradR n}\right) \ket n$, which
is.  The effect on the nuclear Schr\"odinger equation is to
redefine the momentum operator, \begin{equation}\P\psi \defn
(-i\hbar\gradR + \hbar\A_n)\psi,\label{Pop}\end{equation} Here
$\A_n(\R)$ is given by \eqref{vecpot}. Thus, the field whose line
integral gives the electronic geometric phase appears as a vector
potential in the nuclear Hamiltonian.  There it corresponds to an
effective magnectic field $\V_n = \gradR \times \A_n$ which
modifies the nuclear dynamics.  This is called {\it geometric
magnetism}.

The situation described above is quite general.  Whenever there are
two coupled systems evolving on different time scales, we can
effect an approximate description of the slow system by averaging
over the fast motion.  There appears in the slow dynamics an effective
vector potential and associated magnetic field.  The vector
potential and magnetic field are precisely the fields whose line
integrals and fluxes describe the geometric phase.  (In cases where the
coupling involves not only the slow coordinates but the
slow momenta as well,
this description has to be modified somewhat.)

Geometric magnetism is responsible for a variety of phenomena. Mead
and Truhlar analysed its effects in nuclear rotation-vibration
spectra, which have since been observed experimentally (cf
Section~\ref{pseudo}).  These effects can be nontrivial even when
time-reversal symmetry is present (cf Section~\ref{bas.tr}), due to
the change in sign of the electronic wavefunction $\phi_n(\r,\R)$ when
the nuclear coordinates $\R$ are carried round a degenerate
configuration.  Like the magnetic field of the Aharonov-Bohm effect
(Section~\ref{bas.geophase}), the flux field $\V_n(\R)$ in this case
vanishes everywhere except at these degenerate configurations, where
it is singular. Mead (1980) has called this the ``molecular
Aharonov-Bohm effect''.

Geometric magnetism survives in the classical limit.  When this limit is
applied to the slow system alone, it leads to geometric phase
modifications of the semiclassical (Bohr-Sommerfield) quantization
conditions. These are discussed by Kuratsuji and Iida (1985), Wilkinson (1984),
and Littlejohn and Flynn (1991).

The situation where both the fast and slow systems are fully classical
is discussed in Section~\ref{hannay.react}. Here let us mention a
simple but compelling illustration discussed by Li and Mead (1992).
Let us consider the Born-Oppenheimer approximation applied to an atom,
rather than the molecule.  The electronic energy levels $E_n(\R)$
depend on the atom's centre-of-mass $\R$, and appear as potentials in
the equations of motion $\ddot \R = -\gradR E_n + \F(\R)$, along with
the external force $\F(\R)$.  In the absence of external potentials,
the energy levels are independent of $\R$, and $\F \equiv 0$; the
centre-of-mass behaves as a free particle.  Now suppose the atom moves
in a uniform magnetic field.  The energy levels are still
$\R$-independent (by translation invariance), but there now appears in
the centre-of-mass dynamics the Lorentz force $\F = +Ze\dot \R \times
\B$.  The predictions of this analysis are spectacularly unphysical;
walking through the earth's magnetic field, you would experience a
vertical force sufficient to hurl you into the air.  What is missing
from this analysis is geometric magnetism from the electrons; the field $\V_n(\R)$
introduces into the centre-of-mass dynamics a second Lorentz force
which either exactly or nearly (in case the electron eigenstate has a
nonzero magnetic moment) cancels the first.

\section{Theoretical Developments\label{theory}}
A number of theoretical developments of the geometrical phase are
described.  These include its mathematical setting in terms of
holonomy under parallel transport, as well as several generalizations
which remove the assumptions of nondegeneracy and adiabaticity.

\subsection{Geometric phase as holonomy\label{theory.hol}}
Objects rigidly transported round a closed path in a curved space can
return with a different orientation -- this is {\it holonomy}.  Here
is a simple (and surely physical) example. Hold your right arm
outstretched before you with your palm facing down.  Keeping your wrist rigid,
swing your arm into your chest, then raise it clockwise in front of
you, and finally swing it away downwards until it is outstretched before you
again.  Your arm has returned to its original position, but your hand
has turned clockwise through $90^\circ$.

The mathematical setting for this phenomenon is the theory of
connections on fibre bundles.   What follows in Sections~\ref{theory.hol.fb}--
\ref{theory.hol.chern} is an informal account of the relation between
this theory and the geometric phase.  No attempt is made at
mathematical precision; the intention is to convey a
sense of the principal ideas involved.
Some standard
references are (Eguchi et al 1980) and (Nash and Sen 1983).

\subsubsection{Fibre bundles\label{theory.hol.fb}}
A {\it fibre bundle} $E$ is a space, or
manifold, composed of two smaller ones, the {\it base manifold} $B$
and the {\it fibre} $F$.  The fibre bundle is constructed by attaching
a copy of the fibre to each point of the base in a particular way.  The simplest
case, a {\it trivial} fibre bundle, is where $E$ is just the cartesian
product $B\times F$ of the base with the fibre.
A general fibre bundle cannot be so expressed.  While it
is always possible to partition it into smaller pieces which are
themselves cartesian products $M_\alpha\times F$ of subsets of the base
with the fibre,
these pieces may fit together in an interesting and
nontrivial way, which reflects the topological properties of the bundle.

The configuration space $E$ of an outstretched arm can be described as
a fibre bundle.  In this case, the base manifold is the unit sphere,
whose polar coordinates $(\theta,\phi)$ determine the arm's direction.
The fibre is the unit circle, whose single angle coordinate $\chi$
describes the orientation of the hand relative to some reference
position.
In this case $E$ is not simply the cartesian product $M\times F$.
This is because it is not possible to assign, in a continuous way, a
reference hand position for every arm direction.  This is a consequence
of the ``hairy ball theorem'', ie the fact that it is impossible to comb
flat a hairy ball without introducing singularities (eg crowns,
cowlicks, bald patches).

As in the arm example, it is often the case in physical applications
that the fibre describes internal degrees of freedom.  Usually the fibre
has some additional mathematical structure as well.   If it is a real or
complex vector space (describing spin degrees of freedom, for example),
then $E$ is called a {\it real} or {\it complex vector bundle}.   If the
vector space is one-dimensional, $E$ is called a {\it line bundle} (real
or complex).  Or else, the fibre might be a group (for example, the two-
or three-dimensional rotation group,  describing the axial or
spatial orientation of a rigid body).  Then $E$ is called a {\it
principal bundle}.

\subsubsection{Connections and curvature\label{theory.hol.con}} To a
given path $P$ on the base there correspond innumerable paths
in the
fibre bundle
which project down to it; a motion in the base leaves the
fibre motion undetermined. A {\it connection} is a rule for
associating to $P$ a particular path $\hat P$ through a given
point in the bundle, the
lifted path (see Fig.~1).
A point moving along the lifted path
$\hat P$ is said to be {\it parallel transported}.  The
connection rule is expressed in differential form by making the
fibre velocity
a linear function of the base velocity, which can
also depend on the base and the fibre positions.  For vector
bundles, the dependence on fibre position should be linear as well.
There are analogous restrictions  in the case of principal bundles. In
physical applications, a particular connection is often suggested by
kinematical or dynamical considerations, eg, moving your arm without
unnecessarily twisting your wrist.

Let us consider the lift $\hat C$ of a closed path $C$ on the base (see
Fig.~2).  Its endpoints necessarily lie in the same fibre, but
they need not coincide.  If they do not, the difference between the
endpoints is called the {\it holonomy} of $C$.  In the outstretched arm
example, $C$ is the spherical right triangle described by your arm, and
its holonomy is the $\pi/2$ turn of your hand which results from
traversing it in the manner prescribed. In case there is holonomy for
arbitrarily small closed paths, the connection is said to have {\it
curvature}.  Curvature describes  the infinitesimal holonomy acquired
round an infinitesimal closed paths, and may be expressed in terms of
derivatives (a generalized curl) of the connection.

Let us consider a very simple example. We
take the base $M$ to be  three-dimensional Euclidean space $\{\R\in
\Rn^3\}$, the fibre $F$ to be one-dimensional complex vector space $\{z\in
\Cn\}$, and  the bundle $E$ to be their cartesian product $\{(\R,z)\in
\Rn^3\times \Cn\}$.  Thus $E$ is a trivial complex line bundle. A connection
specifies the fibre velocity $\dot z$ as a function of the fibre
position $z$, $\R$ and $\dot \R$ whose
dependence on $z$ and $\dot \R$ is linear.  This relation may be
expressed as $\dot z =  (\W(\R)\cdot \dot\R)\,z$, where $\W(\R)$ is a complex
vector field.  $\W(\R)$ is called the connection form, or  simply the
connection.

Given a motion $\R(t)$ in the base, its lifted motion in the fibre
is  given by $z(t) = \exp\left(\int_{\R(t)} \W(\R)\cdot d\R\right)
z_0$, where $z_0$ denotes the initial position in the fibre of the
lifted path. If $\R(t)$ describes a closed path, then its holonomy
is the scalar factor $\exp \Gamma$, where $\Gamma = \oint_{\R(t)}
\W(\R)\cdot d\R$.  Using Stokes' theorem, we may express $\Gamma$
as the flux of a vector field $\V = \gradR\times \W$ through a
surface $S$ bounded by $C$.  $\V$ is the curvature of the
connection $\W$.

Of particular relevance to the geometric phase is the case in which
the connection preserves the length $|z(t)|$ of vectors along the
lifted path.  Then the connection is said to be {\it hermitian}.  The
connection form $\W(\R)$ is purely imaginary, and may be written as
$i\A(\R)$, where $\A(\R)$ is real.

\subsubsection{The geometric phase revisited\label{theory.hol.geo}}
After seeing a preliminary version of Berry's 1984 paper, B~Simon (1983)
observed that the geometric phase finds its natural mathematical
expression as the holonomy of a connection on a fibre bundle.  The fibre
bundles in question are the complex line bundles $E^{(n)}$ associated with
nondegenerate  eigenstates of a quantum Hamiltonian $H(\R)$. The base
manifold is the parameter space (we continue to assume this to be
$\Rn^3$), and the fibres are vectors $\ket{\psi}$ in Hilbert space which
satisfy the Schr\"odinger equation  $(H(\R) - E_n(\R))\ket{\psi} = 0$;
these are just the unnormalized $n$th eigenstates, and are determined up
to a complex scalar.

Given an eigenstate $\ket {\psi}$ of $H(\R)$, a connection
determines an eigenstate $\ket {\psi} + \ket {d\psi}$ of the
nearby Hamiltonian $H(\R + d\R)$ by fixing its norm and phase.  On
physical grounds it is sensible to require that the connection
conserve probability, ie that $\ket {\psi} + \ket {d\psi}$ and
$\ket {\psi}$ both have unit norm. This implies that $\Re \braket
{\psi} {d\psi} = 0$.  The simplest rule for fixing the phase of
$\ket{\psi} + \ket{d\psi}$ is to take $\Im \braket {\psi} {d\psi}
= 0$ as well.  This determines completely a connection on
$E^{(n)}$, which is defined by the relation
\begin{equation}\braket {\psi} {d\psi} = 0.\label{Simon}\end{equation} Under
parallel transport round a closed path $C$, an eigenstate $\ket
{\psi}$ acquires a phase factor $\exp (i\gamma_n)$.  This phase
factor, the holonomy of $C$, is precisely the geometric phase
factor.  This follows by noting that the connection \eqref{Simon}
is just the condition \eqref{parallel} derived from adiabatic
evolution.

There is an exact mathematical analogy between the outstretched arm and
a variant of the spin example of Section~\ref{bas.spin}.  If a spin-1
(instead of a spin-1/2) particle, initially in its down-state, is  driven by
a magnetic field which is slowly cycled round a closed path $C$, then its
geometric phase $\gamma_1$ is precisely the turn in the hand of an arm
describing the same circuit.  Both are equal to the solid angle
subtended by $C$.

\subsubsection{Chern classes\label{theory.hol.chern}}
Some fibre bundles are trivial cartesian products $M\times F$, and others
are not.  Topology lies at the root of this distinction, and
{\it characteristic classes} make this distinction precise.  They describe
an intrinsic twistedness in fibre bundles which cannot be combed
away.  Bundles whose characteristic classes are the same can be
smoothly transformed into one another; bundles whose characteristic
classes are different cannot.

For complex vector bundles,  the characteristic classes are called {\it
Chern classes}.  The number of Chern classes is equal to the dimension
of the fibre.  The $r$th Chern class associates an integer $c_r(N)$, the
{\it Chern number}, to every $2r$-dimensional, boundaryless submanifold
$N$ of the base manifold $M$.  Thus, the first Chern class assigns an
integer $c_1(S)$ to every closed surface $S$ in $M$.  There are
various ways to calculate Chern numbers.  Among these, there are
formulas in terms of integrals over $N$ involving the curvature.

An illustrative example is the celebrated Gauss-Bonnet theorem from
classical differential geometry.  The theorem states that the integral
$\int_S K dS$ of the Gaussian curvature $K$ over a closed one-sided
surface $S$ is equal to $2\pi$ times an integer $\chi$, the Euler
characteristic.  The Euler characteristic is a topological feature of
the surface. It is related by the  formula $\chi = 2 - 2g$  to the genus $g$,
the number of handles on $S$ when it is regarded through a
topologist's eyes as a many-handled sphere. The Gauss-Bonnet theorem can
be verified immediately  for the unit sphere ($K = 1$,\ $g = 0$).  What
is remarkable is the implication  that $\int_S K dS = 4\pi$ for any surface
smoothly deformable to a sphere, but $\int_S K dS \ne 4\pi$ for any surface
(such as a torus) which is not.

The Gauss-Bonnet theorem can be formulated in terms of fibre bundles.
The  bundle consists of the surface $S$ (the base) together with
its tangent vectors at each of its points (the fibres).  A connection is
defined through the Riemannian rule for parallel transport on a curved
surface. By identifying two-dimensional real tangent vectors with
one-dimensional complex ones, the fibre bundle may be regarded as a
complex line bundle, and its first Chern number $c_1(S)$ turns out to be
twice the integral $\int_S K dS$ of the Gaussian curvature.

Returning to considerations of the geometric phase, let us
consider the complex line bundles $E^{(n)}$ of quantum
eigenstates, discussed in the preceeding section.  Each has a
single Chern class $c^{(n)}_1$ which assigns an integer to every
closed surface $S$ in parameter space.  This integer is given by
\begin{equation}c^{(n)}_1(S) = \oint_S \V_n\cdot d\S = \int_B \gradR\cdot\V_n
\,d\R,\label{Chern}\end{equation} where in the second equation we
have used the divergence theorem; $B$ is  a three-dimensional
region bounded by $S$.  From \eqref{degen}, the last integral is
precisely the number of energy level degeneracies (points $\R_*\in
B$ where $E_n(\R_*) = E_{n\pm1}$) contained in $S$, counted with
their appropriate sign (see Section~\ref{bas.degen}). In terms of
the electromagnetic analogy (Section~\ref{bas.geophase}), it is
the number of magnetic monopoles in $S$ counted with their charge.
Thus, energy level degeneracies manifest themselves in the
nontrivial topology of the eigenstate line bundles $E^{(N)}$.

\subsection{Degenerate eigenstates and
nonabelian geometric phases\label{theory.nonab}} If a system possesses
a sufficient degree of symmetry, its energy levels will be degenerate.
Two examples are the angular momentum multiplets of rotationally
symmetric systems, and Kramers degeneracy in
systems with time-reversal symmetry and an odd number of fermions.  If
the symmetry persists throughout the parameterized family $H(\R)$,
then so do these degeneracies.  Under a slowly varying Hamiltonian
$H(\R(t))$, the adiabatic theorem implies that transitions to states
outside a degenerate multiplet are small.  However, transitions to
states within the multiplet need not be, no matter how slow the
variation (transitions between degenerate states cost no energy).

F Wilczek and A Zee (1984) generalized Berry's analysis to
degenerate eigenstates.  They showed that under adiabatic cycling
round a closed circuit $C$, such states need not return to
themselves up to a phase, but instead can evolve into new states
within the multiplet.  The transformation from initial to final
states is generated by a unitary matrix $U(C)$, called the {\it
nonabelian geometric phase}.  Nonabelian geometric phases can be
interpreted as holonomy, but of a more general sort than was
considered in Section~\ref{theory.hol.geo}. There is a
corresponding generalization of geometric magnetism.

The analysis, which runs parallel to the nondegenerate case,
proceeds  as follows. Let $E(\R)$ denote an $r$-fold degenerate
energy level whose degenerate eigenspace is spanned by an
orthonormal basis $\ket{j(\R)}, j = 1,2,\ldots,r$.  We consider
evolution under a slowly changing Hamiltonian $H(\R(t))$ of states
$\ket{\psi_j(t)}$ initially equal to $\ket{j(\R(0))}$. Neglecting
transitions to states outside the multiplet, we look for
approximate solutions of the form \begin{equation}\ket{\psi_j(t)}
= \exp\left(-i/\hbar\int_0^t E(\R(t))d\tau\right) \sum_{k = 1}^r
U_{jk}(t) \ket{k(\R(t))}, \label{ansatz3}\end{equation} ie, a
superposition of evolving degenerate eigenstates multiplied by an
overall dynamical phase factor. For the inner products $\braket
{\psi_k(t)} {\psi_j(t)}$ to be conserved in time, the $r\times r$
matrix $U(t)$ must be unitary.  Its evolution is determined by
requiring $\me {k(\R)} {(i\hbar \partial_t - H)} {\psi_j(t)}$ to
vanish for each $k$ (it would automatically if ${\psi_j(t)}$ were
an exact solution of the Schr\"odinger equation). This leads to
the equation of motion \begin{equation}\dot U(t) = -i U(t) \Amb
(\R(t))\cdot \dot\R = -iU(t) \Am(t), \label{Udot}\end{equation}
where $\Amb(\R) = (\Am^{(x)}, \Am^{(y)}, \Am^{(z)})(\R)$ is a
vector of $r\times r$ Hermitian matrices whose matrix elements are
given by \begin{equation} \A^{(\alpha)}_{jk}(\R) = -i \braket
{k(\R)} {\nabla_\alpha j(\R)},\label{nonabA}\end{equation} and
$\Am(t) = \Amb(\R(t))\cdot \dot\R(t)$.

The solution of \eqref{Udot} is given by the time-ordered product
\begin{multline} U(t)={\rm T} \exp \left( -i\int_0^t \Am(\tau)\,d\tau\right)\\
 = \Id + \int_{0<\tau<t} \Am(\tau)\,d\tau +
\int\int_{0<\tau<\tau'<t} \Am(\tau)\Am(\tau')\,d\tau d\tau'
+ \cdots,\label{Tord}\end{multline}
where $\Id$ is the identity matrix.  The time ordering indicates that
in expanding the exponential of the integral, all matrices are to be ordered
with later times to the right.  If the matrices $\Am(\tau)$ at
different times commute with each other,
then the ordering of the factors is irrelevant,
and the r.h.s.~of \eqref{Tord} simplifies to $\Id + \int_{0<\tau<t}
\Am(\tau)\,d\tau + (\int_{0<\tau<t} \Am(\tau)\,d\tau)^2/2! + \cdots =
\exp\left(-i\int_0^t \Am(\tau)\,
d\tau\right)$, eg the ordinary exponeniated integral.  In particular,
if $r = 1$, then $\Am(\tau)$ is a scalar, and $U(t)$ is simply a phase
factor.

Suppose the Hamiltonian is taken round a closed cycle $C$. Then
the relation between initial and final states is described, apart
from the dynamical phase factor, by the unitary matrix
\begin{equation}U(C) = {\rm T} \exp \left( -i\oint_C \Amb(\R)\cdot
d\R\right). \label{nonabhol}\end{equation} $U(C)$ is the
nonabelian geometric phase. The terminology derives from the fact
that the factors $\Amb(\R)\cdot d\R$ in the time-ordered product
do not in general commute.  They do, however, in the nondegenerate
case $r=1$; then $U(C)$ is just the ordinary (abelian) geometric
phase factor $\exp (i\gamma(C))$. Like the ordinary geometric
phase, $U(C)$ depends only on the path $C$ and not on the rate at
which it is traversed.

For infinitesimal circuits, $U(C)$ is of the form $\Id +
i\Gamma(C)$, where $\Gamma(C)$ is an infinitesimal Hermitian
matrix given by the flux of the covariant curl of $\Amb$,
\begin{equation}\Vmb = \gradR \times \Amb - \Amb\times \Amb,
\label{nonabV}\end{equation} through the infinitesimal area
bounded by $C$. (Note that because the components of $\Amb$ are
matrices, the cross product $\Amb\times \Amb$ does not vanish,
just as the cross product $\bsigma\times\bsigma = 2i\bsigma$ of
Pauli matrices does not vanish.)  This result does not extend to
finite circuits, though, as there is no nonabelian version of
Stokes' theorem.

Like the ordinary geometric phase, $U(C)$ may be
interpreted as holonomy, but now on a complex $r$-dimensional vector
bundle rather than a line bundle.  The base manifold is parameter
space ($\Rn^3$ in the examples
considered here), and the fibers are the degenerate eigenspaces
spanned by $\ket{j(\R)}$.  The matrix-valued vector field $\Amb(\R)$
is the connection form, which describes how states in the fibres are
parallel transported along paths $\R(t)$ in the base.
$\Vmb(\R)$ is the associated curvature form.

The nonabelian geometric phase does not depend on the choice of
basis $\ket{j(\R)}$ in any intrinsic way. Under a unitary change
of basis \begin{equation}\ket{j(\R)} \rightarrow \sum_{k = 1}^r
W_{jk}(\R)\ket{k(\R)}, \label{nonabgt}\end{equation} the
nonabelian version of the gauge transformation \eqref{gt}, $U(C)$
transforms according to $U(C)\rightarrow W U(C) W^{\dag}$.  This
unitary conjugation represents simply the change of basis applied
to $U(C)$ and leaves its observable properties unchanged.  The
fields $\Amb(\R)$ and $\Vmb(\R)$ transform according to
\begin{equation}\Amb \rightarrow W\Amb W^{\dag} - i\gradR W
W^{\dag}, \quad \Vmb \rightarrow W\Vmb
W^{\dag}\label{nonabgt2}\end{equation} under \eqref{nonabgt}.
These transformation properties characterize the potential and
field strength of Yang-Mills field theories, generalizing the
electromagnetic analogy with the geometric phase (cf Section~
\ref{bas.geophase}) to the nonabelian case.

Given a particular closed circuit $C$, one can choose the basis
$\ket{j(\R)}$ so as to diagonalize $U(C)$.  Then under
parallel-transport round $C$, each basis state returns to itself up to
a phase factor.  Under certain conditions, one can find a basis which
diagonalizes every holonomy, regardless of $C$.  In such cases, the
degenerate and nondegenerate cases are essentially the same.  In
general, though, there is no choice of basis which renders every
$U(C)$ diagonal.  What distinguishes the general case is that the
generators of the symmetries responsible for the degeneracies depend
on parameters themselves.  A particular example, the
$\Lambda$-doubling of electron doublets in diatomic molecules, is
discussed by Moody et al.\ (1986).

The discussion of geometric magnetism in the Born-Oppenheimer
approximation (cf~Section~\ref{bas.geomag}) can also be
generalized to the degenerate case.  As before, we regard the
parameters $\R$ as the coordinates of a slow system to which the
original one, the fast system, is coupled.  The Hamiltonian is
given by $P^2/2M + H(\R)$, where $\P$ is the slow momentum and the
dependence on the fast coordinates and momenta is left implicit in
$H(\R)$.  We look for approximate eigenfunctions of the coupled
system in which the fast system belongs to a degenerate multiplet
of $H(\R)$.  These are of the form $\Psi(\r,\R) = \sum_{j=1}^r
\psi_j(\R)\phi_j(\r,\R)$, where $\phi_j(\r,\R) = \braket {\r}
{j(\R)}$ are the fast eigenfunctions (which depend parametrically
on the slow variables $\R$). Substituting this ansatz into the
stationary Schr\"odinger equation and making approximations
similar to those in the nondegenerate case (the details are
omitted), one sees that the slow system is described by a
multicomponent wavefunction $\vpsi(\R) =
(\psi_1(\R),\ldots,\psi_r(\R))$ which is an eigenfunction of the
matrix-valued Hamiltonian \begin{equation}\Hm = (\P\Id -
\Amb)^2/2M + E(\R)\,\Id.\label{nonabH}\end{equation} Here $\A(\R)$
is the matrix-valued vector potential of \eqref{nonabA} and $\Id$
is the identity matrix.  Thus, coupling to a degenerate eigenstate
introduces a generalized form of geometric magnetism, in which the
slow Hamiltonian acquires spin degrees of freedom which are
coupled to a Yang-Mills-like potential.

Finally, there is a connection between the nonabelian geometric phase
higher-order degeneracies, ie points in parameter space where two or
more degenerate levels coalesce.  These points act as
generalized monopoles sources for the
nonabelian field $\Vmb$.

\subsection{Nonadiabatic evolution\label{theory.nonad}}
Geometric phases appear in contexts other than adiabatic evolution.
An important generalization was found by Y Aharonov and J Anandan (1987), who
showed that geometric phases accompany any cyclic evolution, not
necessarily adiabatic, of a quantum state.  Subsequently, J Samuel and R
Bhandari (1988) showed that geometric phases can arise in evolutions which are
nonunitary, not cyclic, and even discontinuous.  What underlies these
generalizations is a natural prescription for the parallel transport of
quantum phase factors.

The setting for the Aharonov-Anandan phase is quite general.  A
state $\ket{\psi(t)}$ evolves according to the Schr\"odinger
equation $(i\hbar\partial_t - H(t))\ket{\psi} = 0$ under a
(usually) time-dependent Hamiltonian $H(t)$.  After a time $T$ the
state is presumed to return to itself up to a phase factor, so
that $\ket{\psi(T)} = \exp(i\beta)\ket{\psi(0)}$.  In the trivial
case where $H$ is time-independent and $\ket{\psi}$ is an
eigenstate, $\beta$ is simply the dynamical phase $-ET/\hbar$.
Guided by this simple case, we are led to subtract from $\beta$ a
generalized dynamical phase, namely the time integral of the
expectation value of energy \begin{equation}\beta_d =-
\frac{1}{\hbar} \int_0^T \me {\psi} {H} {\psi}\,dt = \Im \int_0^T
\braket {\psi} {\dot \psi}\,dt.\label{AAd}\end{equation} The phase
which remains, $\beta_g \defn \beta - \beta_d$, is given by
\begin{equation}\beta_g = \beta - \beta_d = \arg \, \braket {\psi(0)}
{\psi(T)} - \Im \int_0^T \braket {\psi} {\dot \psi}\,dt
\label{AA}\end{equation}

What makes the decomposition $\beta = \beta_d + \beta_g$
interesting is that $\beta_g$ is, in a sense to be explained, a
geometric phase. First, $\beta_g$ depends only on the path through
Hilbert space described by $\ket{\psi(t)}$, and not on the rate
this path is traversed.  Indeed, the state $\ket{\tilde\psi(t)} =
\ket{\psi(\tau(t))}$ describes the same path but, because of the
reparameterization  $t\rightarrow \tau(t)$,  at a different rate.
However, it is easy to verify that its geometric phase $\beta_g$,
as given by \eqref{AA}, is the same as $\ket{\psi(t)}$'s, even
though its dynamical phase is different (and it may no longer
satisfy the Schr\"odinger equation).   Second, $\beta_g$ does not
depend on the overall phase of $\ket{\psi(t)}$. Indeed, if
$\ket{\psi(t)}$ is replaced  by $\ket{\tilde\psi(t)} = \exp
(i\phi(t))\ket{\psi(t)}$ in \eqref{AA}, a simple calculation shows
that $\beta_g$ is unchanged.  Such a substitution can be useful in
evaluating \eqref{AA}. As $\ket{\psi(t)}$ is periodic up to a
phase, the density matrix $\rho_\psi(t) =
\ket{\psi(t)}\bra{\psi(t)}$ is strictly periodic, and over the
time $0 < t <T$ describes a closed curve $C$ in the space of
density matrices. What the preceding discussion shows is that
$\beta_g$ is a geometric property of $C$.  It depends not on the
rate with which $C$ is traversed, and it can be computed with
\eqref{AA}, substituting any state $\ket{\tilde \psi(t)}$ whose
density $\ket{\tilde \psi(t)}\bra{\tilde \psi(t)}$ is equal to
$\rho_\psi(t)$.

The geometric phase for adiabatic cycles can be obtained as a
special case of \eqref{AA}. Within the adiabatic approximation, a
solution to the Schr\"odinger equation for a slowly cycled
Hamiltonian $H(\R(t))$ is given by $\ket{\psi(t)} = \exp
(i\beta(t))\ket{n(\R(t))}$ (cf \eqref{ansatz2}).  From
\eqref{AAd}, the dynamical phase $\beta_d$ is given by
$-1/\hbar\int_0^T E_n(\R(t))dt$. The geometric phase is easily
computed from \eqref{AA} if  $\ket{\tilde\psi(t)} =
\ket{n(\R(t))}$ is substituted for $\ket{\psi(t)}$ (this is
permissible, since the two differ by an overall phase).  Since
$\arg\braket {n(\R(0)} {n(\R(T))} = 0$, $\beta_g$ is given by
$-\Im \int_0^T \braket {\tilde \psi} {\dot{\tilde \psi}}dt = -\Im
\oint \braket {n(\R)} {\gradR n(\R)}\cdot d\R$, in agreement with
\eqref{geo}.

What underlies this general setting for the geometric phase is
once more the mathematics of connections and fibre bundles.
However, now all of Hilbert space is regarded as a fibre bundle!
The base manifold $\ph$ consists of all pure state density
matrices $\rho$, normalized so that $\Tr \rho = 1$.  (This is
sometimes called the {\it projective Hilbert space}.)  The fibre
attached to $\rho$ consists of all state vectors $\ket{\psi}$
whose density matrices $\ket{\psi}\bra{\psi}$ are equal to $\rho$.
(Thus states in the same fibre differ only by a phase factor.)  To
an infinitesimal displacement $d\rho$ there corresponds
displacements $\ket{d\psi}$ in the fibre satisfying $\ket{\psi +
d\psi}\bra{\psi + d\psi} = \rho + d\rho$. With a little algebra
this reduces to $\ket{d \psi} - \braket {\psi} {d\psi} \ket{\psi}
= d\rho$.  This equation does not determine $\ket{d\psi}$
uniquely; to any solution can be added infinitesimal imaginary
component $id\alpha\ket{\psi}$ along $\ket{\psi}$.  A connection
fixes this component and determines a particular $\ket{d\psi}$ for
a given $d\rho$.  A natural choice is to require that
\begin{equation}\braket {\psi} {d\psi} = 0.\label{HScon}\end{equation} This connection
is precisely the condition \eqref{parallel} found by Berry for
adiabatic cycles, but now generalized to arbitrary cycles in
Hilbert space. Its holonomy is precisely the geometric phase
$\beta_g$ \eqref{AA}.

This setting admits further generalizations.  For example, one can calculate
a geometric phase for a nonunitary evolution $\ket{\phi(t)}$,
for which $\braket {\phi} {\phi}$ is no longer constant,
simply by evaluating \eqref{AA} with
$\ket{\psi(t)} \defn \ket{\phi(t)}/[\braket {\phi} {\phi}]^{1/2}$.

In fact, one can calculate a geometric phase for an open path
$\ket{\psi(t)}$, where $\ket{\psi(T)}$ and $\ket{\psi(0)}$ are no
longer proportional. The formula \eqref{AA} for $\beta_g$ can be
computed for open paths, and the result does not depend on the
parameterization in time of $\ket{\psi(t)}$, and remains unchanged
by the change of phase $\ket{\psi(t)}\rightarrow \exp
(i\phi(t))\ket{\psi(t)}$. Further insight into the interpretation
of the geometric phase for open paths is given in the following
Section~\ref{theory.metric}.

There is also a discrete version of the formula \eqref{AA}. Given
a sequence of states $\ket{\psi_1}, \ket{\psi_2}, \ldots
\ket{\psi_N}$, one can associate a phase according to the formula
\begin{equation}\beta_g = \arg\braket {\psi_1} {\psi_2} \braket {\psi_2}
{\psi_3} \ldots \braket {\psi_N}
{\psi_1}.\label{pan1}\end{equation} One sees easily that $\beta_g$
does not depend on the overall phases of the states
$\ket{\psi_j}$.  If there are just two states, then $\beta_g =
\arg\, |\braket {\psi_1} {\psi_2}| = 0$ identically. But if $N>2$,
the phase $\beta_g$ is typically nonvanishing.  Note that
$\ket{\psi_N}$ need not be proportional to $\ket{\psi_1}$ -- the
sequence of states need not close on itself.

The formula \eqref{pan1} is the discrete analogue of \eqref{AA}, and
defines a geometric phase for discontinuous evolutions.  Thus, a
geometric phase can be defined for processes which include
measurements and wavefunction collapse.  A version of formula
\eqref{pan1} appears in work of Pancharatnam in optics, in which
$\ket{\psi_j}$ represent polarization states.  Panchartnam's
contributions are discussed in Section~\ref{optics.panch}.

\subsection{Geometric metric tensor\label{theory.metric}}
How close are two quantum states $\ket{\psi}$ and $\ket{\phi}$?  The usual
quantum mechanical measure of distance is the Hilbert space norm
$||\psi - \phi||$ of their difference.  There is another,
first introduced by J Provost and G Vallee (1980), namely
the quantity
$1 - |\braket {\psi} {\phi}|^2/\braket {\psi} {\psi}
\braket {\phi} {\phi}$.
Clearly this vanishes if the states are the same.
What distinguishes it from the Hilbert space norm is that
it is unchanged if  $\ket{\phi}$ or $\ket{\phi}$ are multiplied by
a complex scalar. In fact, what this quantity
really describes is the distance between their
normalized density
matrices $\rho_{\psi} = \ket {\psi} \bra {\psi}/\braket {\psi} {\psi}$
and $\rho_{\phi} = \ket {\phi} \bra {\phi}/\braket {\phi} {\phi}$,
and indeed can be alternatively expressed as
$1 - \Tr \rho_\psi \rho_\phi$.
For infinitesimally displaced states $\ket{\phi} = \ket{\psi +
d\psi}$, a little algebra gives
\begin{equation}  ds^2(d\psi) \defn
\frac {\braket {d\psi}{d\psi}} {\braket {\psi} {\psi}} - \frac
{\braket {d\psi} {\psi} \braket {\psi} {d\psi}} {{\braket{\psi}
{\psi}}^2} = \frac {  \me {d\psi} {\Id - \rho_\psi} {d\psi}}
{\braket {\psi} {\psi}} =
-\frac12\Tr\left(d\rho_\psi\right)^2.\label{metric}\end{equation}
This quantity defines a metric on the space of normalized density
matrices $\ph$ (or equivalently, the projective Hilbert space).

The quantum metric enters into the analysis of the
geometric phase in a number of interesting ways.
Below we describe two aspects: a deeper understanding
of the geometric phase for open paths, and physically significant
corrections to the Born-Oppenheimer approximation.

The quantum metric determines geodesics, ie curves $\rho(s)$ which
minimize, or at least render stationary, the distance between two
given density matrices in $\ph$.  These are most easily
described in terms of their corresponding state
vectors; it may be shown that if $\ket{\psi(s)}$ satisfies $d^2\ket{\psi}/ds^2
+ \ket{\psi} = 0$ and if $\ket {\psi(0)}$, $\ket {\psi'(0)}$ are normalized
and orthogonal to one another, then the curve $\rho_{\psi}(s) = \ket{\psi(s)}
\bra{\psi(s)}$ is a geodesic with respect to the metric \eqref{metric},
and the parameter $s$ along the curve is
its arclength.  (The solutions $\ket{\psi(s)}$ are of the form $\cos s\,
\ket{\psi(0)} + \sin s\, \ket{\psi'(0)}$.)

A simple illustration is provided by two-state systems.  In this case,
$\ph$ may be identified with the unit sphere (in the context of
polarization optics, this is the Poincar\'e sphere -- see
Section~\ref{optics.panch}); each density matrix $\rho$ may be
expressed in the form $(I + \ehat.\pauli)/2$, where $\ehat$ is a unit
vector and $\pauli$ are the Pauli matrices.  The quantum metric
\eqref{metric} reduces to the usual metric on the sphere
(up to a scalar factor), and the geodesics
are the arcs of great circles.

As shown in the previous Section~\ref{theory.nonad}, geometric phases
can be defined via \eqref{AA} for open as well as closed paths in $\ph$.
The two cases may be related as follows.  An open path $P$ may be
closed by joining its endpoints with a geodesic. The resulting closed
path, it turns out, has the same geometric phase as $P$.  The
geometric phase for continuous and discontinuous evolutions, \eqref{AA}
and \eqref{pan1}, may be similarly related; given a closed sequence of density
matrices $\rho_1$, $\rho_2, \ldots \rho_N,\rho_1$, we can construct a
closed polygonal path $\rho(s)$ in $\ph$ by connecting its points with
geodesics.  $\rho(s)$
turns out to have the same geometric phase as the discrete sequence $\rho_j$.

The quantum metric also plays a role in the Born-Oppenheimer
description of the  dynamics of a slow system coupled to a
fast one (cf Section~\ref{bas.geomag}).  There it appears in the slow
Hamiltonian as an additional scalar potential $\Phi_n(\R) = \sum_\alpha
(ds^2(\ket{\nabla_\alpha n(\R)})/2M$, which produces a `geometric electric'
force.
The scalar potential appears at a higher order of adiabatic
expansion than the vector potential $\A_n(\R)$; with appropriate
scalings, the geometric magnetic force $\dot \R \times \V_n$ is of the
order of the square root $\sqrt{m/M}$ of the mass ratio of the fast
and slow systems, whereas the ``geometric electric force''
$-\gradR\Phi$ is of the order  $m/M$.
However, the electric force can become significant in the vicinity of
degeneracies $\R_*$ (ie, points where the fast energy level $E_n(\R)$
becomes degenerate).  There the force is repulsive, and so tends to
steer the slow dynamics away from points where the Born-Oppenheimer
approximation would break down.  A discussion of the scalar potential
can be found in the reviews of Berry (1989) and Jackiw (1988).
Additional terms in the slow dynamics of comparable order to geometric
electricity have been analyzed by Weigert and Littlejohn (1993).

\subsection{Beyond adiabaticity\label{theory.beyad}}
\subsubsection{Corrections to the geometric phase\label{corr}}
For the adiabatically cycled eigenstates described by
Eq.~\eqref{ansatz2}, the dynamical and geometric phases are but the
first two terms in an asymptotic expansion for the total phase.  An
elegant iterative scheme for evaluating higher order corrections was
developed by MV Berry (1987c).  These corrections to
the geometric phase turn out to be geometric phases
themselves.  The full asymptotic series diverges in a manner
which reveals in a fundamental way the breakdown of the
adiabatic approximation.

The method is illustrated by the example of a spin-1/2 particle in a
magnetic field (cf Section~\ref{bas.spin}).  We consider a
time-dependent Hamiltonian $H_0(\tau)=\frac12 \mu\B_0(\tau)\cdot\bsigma$,
cyclic over the infinite interval $[-\infty,\infty]$ with
$\B_0(\pm\infty) = B_*\zhat$.  $\B_0(\tau)$ is analytic, and we assume
that $B_0(\tau)\ge B_{\text{min}}$ for all $\tau$, so that the Hamiltonian does
not pass through any degeneracies.   $\tau = \epsilon
t$ is a dimensionless time, and $\epsilon$  a small parameter
(with dimensions of frequency) describing the slow rate of change of
the Hamiltonian.  (The corresponding dimensionless small parameter is
$\delta = \epsilon\hbar/\mu B_{\text{min}}$.)

Let us consider the evolution of a spin
state $\ket{\psi(t)}$ with initial condition $\ket{\psi(-\infty)} =
\ket{-}$.
The analysis is facilitated by transforming to a rotating frame in which
the direction $\bhat_0$
of the magnetic field is fixed (say, for definiteness, along $z$).
Such a frame is described by a rotation matrix $\calR_0(\tau)$ satisfying
    \begin{equation}\bhat_0(\tau) = \calR_0(\tau)\cdot
    \zhat.\label{framerot}\end{equation}
This condition does not determine $\calR_0(\tau)$ uniquely, but
only up to a rotation (applied from the left) about $\bhat_0$.
Differentiating with respect to $\tau$ (this is denoted by a dot
$\dot {}$), we obtain a condition
         \begin{equation}\dot \bhat_0(\tau) =  \bomega_0(\tau) \times
         \bhat_0(\tau)\label{rotate}\end{equation}
on the angular velocity $\bomega_0$ of the frame, where
$\omega_{0i} = \frac12 \sum_{jkl} \epsilon_{ijk}\calR_{0kl} \dot
\calR_{0jl}$ ($\epsilon_{ijk}$ is the Levi-Cevita symbol).
\eqref{rotate} does not determine $\bomega_0$ uniquely, but only
up to its component along $\bhat_0$. We fix this indeterminacy
(and the corresponding indeterminacy in $\calR_0$) by requiring
that
    \begin{equation}\bomega_0\cdot\bhat_0 =
    0.\label{condition}\end{equation}
\eqref{condition} is a condition for parallel transport, analogous
to the requirement of moving your arm without unnecessarily
twisting your wrist (cf Section~\ref{theory.hol}).

In the rotating frame, the Hamiltonian $H_1(\tau)$
is given by $\frac12 \mu \B_1(\tau)\cdot\bsigma$, where
    \begin{equation}\B_1(\tau) = B_0(\tau)\zhat -
    \B_{L1}(\tau)\label{b1}\end{equation}
is the effective magnetic field.  $\B_{L1}(\tau) =
\epsilon\hbar/2\mu\, \calR_0^{-1}(\tau)\cdot\bomega_0(\tau)$ is
the Larmor field induced in the rotating frame (the magnetic
analogue of the Coriolis force).  As the rotation is slow, the
Larmor field is small (of order $\epsilon$), and in the lowest
order adiabatic approximation (cf Sections~\ref{bas.adiabatic},
\ref{bas.geophase}) it is simply neglected.  Then $H_1(\tau)$ is
diagonal, and the time-dependent Schr\"odinger equation is easily
solved.  The geometric phase $\gamma_-(C_0)$, half the solid angle
of the cycle $C_0$ described by $\B_0(\tau)$, appears as a direct
consequence of the parallel transport condition \eqref{condition}
when the solution is transformed back to the fixed frame.

If the Larmor field is not neglected, then $\B_1(\tau)$ may be
regarded as a slowly-varying magnetic field to which the above procedure
may be applied as before.  That is, we transform to a second
rotating frame $\calR_1(\tau)$ in which $\bhat_1(\tau)$ is fixed along
$z$. The frame is uniquely determined by imposing the parallel transport condition
$\bomega_1\cdot\bhat_1 = 0$.  Note that
because $\bhat_1$ differs from $\zhat$ by terms of order
$\epsilon$, the angular velocity $\omega_1$  is of order $\epsilon$.
In this second frame, $H_2(\tau) = \frac12 \mu
\B_2(\tau)\cdot\bsigma$, where the effective magnetic field $\B_2(\tau)$ is
given by $B_1\sigma_z - \B_{L2}(\tau)$, with $\B_{L2}(\tau) = \epsilon
\hbar/2\mu\, \calR^{-1}(\tau)\cdot \bomega_1(\tau)$.  The
new Larmor field $\B_{L2}$ is of order $\epsilon^2$.

It is easy to see that this procedure can be performed iteratively; we
can transform through successive rotating frames $\calR_{n}(\tau)$
to compensate the Larmor fields $\B_{Ln}(\tau)$ induced in the
preceding ones.  The geometric phase and its corrections emerge when
the sequence of rotations is unravelled and the solution
transformed back to the fixed frame.  One obtains a formal
series
   \begin{equation}\gamma_-(C_0) + \sum_{k=1}^\infty
   \gamma_-(C_k),\label{asympt}\end{equation}
in which the corrections to the geometric phase are themselves
geometric phases; $\gamma_-(C_k)$ is half the solid angle of the
cycle $C_k$ described by $\B_k(\tau)$.

It turns out that the series \eqref{asympt} is divergent.  Although
successive cycles $C_k$ initially decrease in extent by a factor of
$\epsilon$,  asymptotically the terms $\gamma_-(C_k)$ behave
as $\epsilon^k k!$, a form characteristic of divergent asymptotic series (Dingle
1973).
An optimal approximation is obtained by truncating
the series at its smallest term $k \sim 1/\epsilon$, for which
$\epsilon^k k! \sim \exp (-1/\epsilon)$ is exponentially small.

That the series \eqref{asympt} diverges can be understood on general
grounds.  If, on the contrary, the series converged, along with the
infinite sequence of rotating frames, than upon transforming back to
the fixed frame, we would obtain an exact solution $\ket{\psi(t)}$
of the Schr\"odinger equation which would begin and end, at $t =
-\infty$ and $t = +\infty$ respectively, in the spin-down state
$\ket{-}$.  This would imply a vanishing transition probability
$\braket {+} {\psi(+\infty)}$.  But it is known that, with the
exception of some special Hamiltonians, the transition probability is
not zero, but instead (for $\B_0(\tau)$ analytic) is exponentially
small, of order $\exp(-1/\epsilon)$ (Dykhne 1962).  These
exponentially small transitions (the subject of the following section)
signal the ultimate failure of the adiabatic approximation, and
cannot be captured by the iterative scheme.  It follows that this
scheme must inevitably diverge.

\subsubsection{The geometric amplitude\label{amplitude}}
Adiabatic evolution is an approximation.  For typical slowly varying
Hamiltonians $H(\tau)$, where $\tau = \epsilon t$, no matter how small
the rate of change $\epsilon$, there are transitions between evolving
eigenstates.  The transition probability vanishes as $\epsilon$ goes
to zero, and if $H$ depends analytically on $\tau$ it does so
exponentially rapidly (faster than any power of $\epsilon$).

In a study of adiabatic evolution in two-state systems, MV Berry (1990b)
found that this transition probability may contain a geometric factor.
The setting is similar to that of the preceding Section~\ref{corr}.  We consider
a two-state Hamiltonian $H(\tau) = \R(\tau)\cdot\bsigma$, where
$\R(\tau)$ is analytic in the dimensionless time $\tau = \epsilon t$ and is nonvanishing
for real $\tau$.   As $\tau\rightarrow\pm\infty$, $\R(\tau)$ approaches limiting values
$\R_{\pm*}$ which need not coincide,
so that $\R(\tau)$ need not be closed.
From \eqref{ansatz2}, the adiabatic solution for the up-state is given by
$\ket{\psi(t)} = \exp (i\delta_+(t) + i\gamma_+(t))\ket{+(\R(\tau))}$,
where the eigenstates $\ket{\pm(\R)}$ are given by \eqref{spins}
and the evolving dynamical and
geometric phases $\delta_+(t)$ and $\gamma_+(t)$ are given explicitly by
    \begin{equation}\delta_{+}(t) =
             -\frac{1}{\epsilon \hbar} \int_0^\tau R(\tau)\,d\tau,\quad
          \gamma_{+}(t) = -\frac12
\int_0^\tau
(1-\cos\theta)\dot\phi\,d\tau.\label{phases}\end{equation} Here
$(R,\theta,\phi)$ denote the polar coordinates of $\R(\tau)$, and
the dot $\dot {}$ denotes the derivative with respect to $\tau$.
We have used \eqref{spinflds} for the vector potential
$\A_{+}(\R)$.

Nonadiabatic transitions may be attributed to roots of $R(\tau)$
in the complex $\tau$-plane.  The dominant contribution
to the transition amplitude comes from the root $\tau_c$
closest to the real axis.
(We assume $\tau_c$ is a simple zero of $R^2(\tau)$,
so that it is a square-root branch point of $R(\tau)$.)  Davis
and Pechukas (1977) showed that the transition amplitude may be
obtained by analytically continuing the adiabatic solution around the branch point
$\tau_c$.  (Under this continuation, the up- and down-states
$\ket{+(\R(\tau))}$ and $\ket{-(\R(\tau))}$ are interchanged.)
The transition probability $P_{+-} = |\braket {-(\R(\infty))} {\psi(\infty)}|^2$
is given by a product of factors
$\exp(-|\Delta_+|) \exp(\Gamma_+)$, whose exponents are
obtained from the analytical continuation
of the dynamical and geometrical phases. Explicitly, we have that
    \begin{equation}\Delta_{+}(t) = \frac{4}{\hbar\epsilon}\Im \int_0^{\tau_c}
R(\tau)\,d\tau,\quad \Gamma_+ = -\frac12
\Im\oint\cos\theta\,d\phi.\label{amps}\end{equation} In the
expression for $\Gamma_+$, the contour is taken round the cut
joining $\tau_c$ and its complex conjugate $\tau_c^*$.

The factor $\exp(\Gamma_+/2)$ is the geometric amplitude.  It depends
only on the cycle followed by $\R(\tau)$, not on the slowness
parameter $\epsilon$ nor on $\hbar$.  It can be shown to vanish if
$\R(\tau)$ lies in a plane, or if $\R(\tau)$ can be
rotated into itself about an axis through the origin.  Thus it
vanishes for the well-known Landau-Zener model $\R(\tau) = (a,a,A\tau)$ (Zener 1930),
for which $\R(\tau)$
describes a vertical line.  A simple example
with nonzero geometric amplitude is $\R(\tau) =
(a\cos(\omega\tau^2),a\sin(\omega\tau^2), A\tau)$.  It describes a helix on the
cylinder of radius $a$ which winds for $\tau<0$ and unwinds for $\tau
> 0$.  For this curve $\Gamma_+$ is given by $-a^2 \omega\, \sgn A /A^2$.

The geometric amplitude was observed by Zwanziger et al (1991) in a
nuclear magnetic resonance experiment (cf Section~\ref{nmr}).  A
time-dependent spin Hamiltonian with the requisite properties was
generated by applying a modulated rf magnetic field to a sample of
spin-1/2 nuclei, and the induced magnetization was observed in a rotating
frame.  The logarithm of the transition probability $P_{+-}$ was plotted as
a function of the slowness parameter $\epsilon$.  Theory predicts that
this plot should be linear (due to the dynamical amplitude) with a
nonzero intercept (due to the geometric amplitude).  This behaviour
was confirmed in the experiment, with good quantitative agreement between the
observed and predicted value of $\Gamma_+$.

\section{Classical Systems\label{classical}}
Following Berry's discovery of the geometric phase, JH Hannay (1985)
found an analogous phenomenon in classical mechanics.
Adiabatically cycled classical oscillators  undergo a shift in
their phase of oscillation.  Part of this shift is simply the time
integral of the frequency. But there appears in addition a purely
geometrical contribution, now known as the Hannay angle.  The Hannay
angle includes and generalizes such well-known phenomena as the
precession of the Foucault pendulum.  In many cases it may be directly
observed.  The Hannay angle emerges as the classical
($\hbar\rightarrow 0$) limit of the Berry phase (in cases
where chaos is absent from the classical dynamics).  In the context of
coupled systems, it leads to classical geometric magnetism in the slow
dynamics.

F Wilczek and A Shapere (1989b-1989d) studied phenomena in classical
kinematics similar to the Hannay angle.  A simple example is the
manner in which falling cats and platform divers change their spatial
orientations through a cycle of internal motions.  Provided their
(conserved) angular momentum is zero, the change in their orientation
does not depend on the rate at which the internal motions are
performed.  Classical geometric phases of this type have been
investigated systematically by J Marsden, R Montgomery and coworkers
(Marsden et al.\ 1991, Marsden 1992)
as part of a general theory of Hamiltonian dynamics
with symmetry.

\subsection{The Hannay Angle\label{hannay}}
\subsubsection{The Classical Adiabatic Theorem\label{hannay.cat}}
Imagine a swinging pendulum slowly shortened to half its original
length.  During this process, is anything conserved, either exactly or
approximately?  This question is addressed by the
classical adiabatic theorem.  While energy is not conserved (work is
being done), there is an approximately conserved quantity, the
classical adiabatic invariant, the {\it action}.

The classical adiabatic theorem is most simply described for systems
of one degree of freedom, so we concentrate on this case.  The
dynamics in the two-dimensional phase plane is described by Hamilton's
equations $\dot q = \partial_p H, \dot p = - \partial_q H$.  The
classical Hamiltonian $H(q,p,\R)$ is taken to depend on parameters
$\R$ (like the quantum Hamiltonians of Section~\ref{basic}).  If these
are held fixed, then energy is conserved, and orbits lie on contours of
$H$.  Let us assume the motion is bounded, so that the
energy contours are closed.  The orbits go round them periodically with
oscillation frequency $\omega = (\partial_E I)^{-1}$, where
    \begin{equation} I(E,\R) = \frac{1}{2\pi} \int_{H(q,p,\R \le E} dq\,dp
           \label{action}\end{equation}
is the action.  Apart from the factor of $2\pi$, it is just the
area enclosed by the orbit of energy $E$. If the parameters vary
in time, then energy is no longer conserved. However, if the
variation is slow compared to the oscillation frequency $\omega$,
then the  action \eqref{action} is nearly conserved, and the
variation of the energy is determined implicitly by
     \begin{equation}I(E(t),\R(t)) = I_0 \defn
     I(E(0),\R(0)).\label{action2}\end{equation}
This is the  classical adiabatic theorem.

In higher dimensions, the classical adiabatic theorem generalizes
provided the dynamics is either {\it integrable} or {\it ergodic}.
Integrable systems are characterized by a high degree of symmetry.
They possess $N$ independent constants of the motion (where $N$ is the
number of degrees of freedom) which are in involution (that is, the
Poisson bracket between any pair of them vanishes).  Typical orbits in
an integrable systems execute periodic or quasiperiodic motion on
$N$-dimensional tori embedded in the $2N$-dimensional phase space, and
the equations of motion may be solved more or less explicitly.  In
sharp contrast stand the ergodic systems, for which energy is the only
constant of the motion.  Typical orbits cover the $(2N-1)$-dimensional
energy surface uniformly, so that time averages and microcanonical
ensemble averages coincide.  The adiabatic theorems in higher
dimensions are rather more difficult to state and considerably harder
to prove.  The difficulty is related to the fact that the adiabatic
invariants are not as well conserved as in the one-dimensional case
(Lochak and Meunier 1988).

\subsubsection{Basic account of the Hannay angle\label{hannay.bas}}
The classical adiabatic theorem \eqref{action2} determines the time
evolution of the energy of a one-dimensional oscillator under a slowly
changing Hamiltonian. J Hannay (1985), motivated by Berry's analysis of
the quantum geometric phase, asked a more refined question: what is the
evolution of the phase of oscillator?

To frame this question precisely we introduce canonical coordinates, the
{\it action-angle variables} $(\theta,I)$.  These are related to $q,p$
by a  canonical transformation $q = q(\theta,I,\R), p = p(\theta,I,\R)$
which depends on parameters. The action $I$, the generalized momentum,
determines the energy of the orbit according to \eqref{action}. The angle
$\theta$, the generalized coordinate, describes the phase of the
oscillator, and varies periodically around the orbit with period $2\pi$.
Expressed in terms of action-angle variables, the Hamiltonian  $H(I,\R)$
is independent of $\theta$, so the canonical equations of motion, $\dot
\theta = \partial_I H, \dot I = -\partial_\theta H$, imply that  $I$ is
constant while $\theta$ evolves at a constant frequency $\omega
\defn \partial_I H$.

Now suppose the parameters $\R$ of the Hamiltonian are taken slowly round a
closed cycle $C$, returning to their original values $\R(0)$ after a time $T$.
Since the action is nearly conserved, a trajectory
returns approximately to the contour of $H(I,\R(0))$ that it started on.  Of
course, it will be displaced by an angle $\Delta \theta$ along the orbit
from its initial position.
Hannay obtained the
two dominant contributions to $\Delta \theta$ (the corrections are of
order 1/T).  The first is the
{\it dynamical angle}
$\theta_d = \int_0^T \omega(I,\R(t)) dt$.
It is simply the time integral of the frequency, taking into account its slow
variation with parameters.  The second, unexpected, contribution,
    \begin{equation}\theta_g(I,C) = -\frac{\partial}{ \partial I} \oint_C
            \A(I,\R)\cdot d\R,\label{theta}\end{equation}
is the {\it geometric angle}, or the {\it Hannay angle}.  It is
expressed in terms of the line integral of a vector field
$\A(I,\R)$ given by
    \begin{equation}\A(I,\R) = \frac{1}{ 2\pi}
             \int_0^{2\pi} p(\theta,I,\R)\gradR q(\theta,I,\R)\,
            d\theta.\label{Ac}\end{equation}
The Hannay angle is geometric in the same sense as the geometric
phase; it depends only on the parameter cycle $C$, and not on the
rate it is traversed. Using Stokes' theorem, the Hannay angle can
also be expressed as the surface integral $-\partial_I \int_S
\V(I,\R)\cdot d\S$, where $S$ is a surface bounded by $C$ and  $\V
= \gradR \times \A$. From \eqref{Ac},
    \begin{equation}\V(I,\R) = \frac{1}{ 2\pi} \int_0^{2\pi} \gradR p(\theta,I,\R)\times
                  \gradR q(\theta,I,\R)\,
            d\theta.\label{Vc}\end{equation}

As with the geometric phase, the presence of time-reversal symmetry (cf
Section~\ref{bas.tr}) imposes constraints on the Hannay angle.  In
particular, if the orbits themselves are time-reversal symmetric (for
all values of parameters), then the Hannay angle vanishes identically.

Like the geometric phase, the Hannay angle may be interpreted as the
holonomy of a connection on a fibre bundle (cf
Section~\ref{theory.hol.geo}).  The parameters $\R$ constitute the
base manifold, and the fibres attached to them are orbits of the
Hamiltonian $H(q,p,\R)$ with given action $I$.  These orbits are
parameterized by the angle $\theta$, and the connection
$\partial_I\A(I,\R)$ describes how $\theta$ changes under parallel
transport.

The Hannay angle can be detected by determining the initial and final
angles of an adiabatically cycled oscillator, and subtracting from their
 difference the dynamical angle $\theta_d$. As $\theta_d$ depends only
on the parameter cycle $\R(t)$, it can be computed without solving the
equations of motion explicitly.  In this way, Hannay's analysis,
together with the adiabatic theorem \eqref{action2}, provides a complete
asymptotic solution of the equations of motion for slowly varying
Hamiltonians in one dimension.   The analysis generalizes directly to
integrable systems in higher dimensions, where there are as many Hannay
angles as there are degrees of freedom.  However, detecting them can be
a more delicate matter, as the adiabatic theorem and corrections thereto
become more complicated in higher dimensions (Golin et al.\ 1989).
For ergodic systems, there is no general analogue of the Hannay angle,
though a related holonomy phenomenon emerges in some special cases
(Robbins 1994, Jarzynski 1995, Schroer 1996).

\subsubsection{Examples\label{hannay.ex}}
A bead moving freely round a planar wire loop  provides a simple example
of the Hannay angle.  Suppose the bead has unit mass, and the loop has
circumference $C$, area $A$, and is otherwise arbitrary in shape.
While the system as a whole has time-reversal symmetry, the
individual orbits do not
(they are either clockwise or counterclockwise).  The action $I$ is just
the velocity $v$, and the angle $\theta =  2\pi s/C$ is proportional to
the distance $s$ along the wire.  Suppose the loop is slowly rotated
once around an interior point over a time $T$, and let $\Delta
s$ denote the total distance the bead has travelled during this time.
The dynamical contribution  $s_d$ (which is $C/2\pi$ times the dynamical angle
$\theta_d$) is just $vT$ (the adiabatic theorem implies $v$ is nearly
constant).  The geometric contribution $s_d$ may be computed to be
$-4\pi A/C^2$.  It  accounts for the distance the bead has
slipped as the wire turned.
For a fixed circumference $C$, $s_d$ assumes its
maximal value, $C$, for a circular loop; in this case, the bead simply slips a
full revolution. For thin, elongated loops, $s_d$
approaches zero; the only slippage occurs at the hairpin turns.

The second example is the family of harmonic oscillators $H = (Z-X)p^2/2
+ Ypq + (Z+X)q^2$, where the term in $qp$ breaks time-reversal symmetry.
For $X^2 + Y^2 < Z^2$, the motion is bounded and oscillatory, and we
consider adiabatic cycles confined to this region of parameter space.
The flux  field $\V$ of \eqref{Vc} may be computed to be
    \begin{equation} \V(I,\R) =
        -\frac{I}{ 2} \frac{\R}{ (Z^2 - X^2 -
        Y^2)^{3/2}}.\label{gsho}\end{equation}
Note that because $\V$ is linear in $I$, the Hannay angle
$\theta_g$ is independent of $I$. $\theta_g$ may be described
geometrically as follows. We regard the parameter space as
three-dimensional Minkowski space, in which $X$ and $Y$ are
space-like coordinates, $Z$ is time-like, and the region $X^2 +
Y^2 < Z^2$ is the interior of the light-cone.  Then  $\theta_g(C)$
is the area of the projection of the closed curve $C$ onto the
unit mass shell $Z^2 - X^2 - Y^2 = 1$.

The precession of a Foucault pendulum is yet another example of a Hannay
angle.  Consider a spherical pendulum initially set to oscillate in a
plane. During the course of a day, the axis of gravity (whose
direction represents
the parameters of the pendulum) is carried once round a line of latitude
$\alpha$.  At the day's end, the plane of oscillation has turned through
an angle $2\pi(1 - \cos \alpha)$.  This is just the solid angle
described by the axis of gravity, and may be computed as a Hannay angle
using \eqref{theta}. In fact,  the phenomenon described by the Hannay
angle is more general.   If one could adiabatically but otherwise
arbitrarily turn the earth around, returning it to its original
orientation,  the plane of oscillation of the spherical pendulum would
still precess through an angle equal to the solid angle subtended by
the earth's axis.  The spherical pendulum is a two-degree
of freedom system, integrable because of its axial symmetry.
Therefore, it has two Hannay angles, one each for its polar and
azimuthal degrees of freedom. For planar oscillations, however, only the
azimuthal Hannay angle, which describes the Foucault precession, is
nonzero.

Let us briefly mention two final examples drawn from plasma physics
and astronomy, respectively.  Littlejohn (1988) found a Hannay angle
in the cyclotron oscillations of charged particles in strong magnetic
fields.  Berry and Morgan (1996) (see also Golin et al.\ (1989)) compute
``the Hannay angle of the world'', an annual displacement of the earth
along its orbit of approximately $-150$ meters (or $-2\times 10^{-4}$
seconds of arc) caused by the adiabatic and periodic variation of
Jupiter's gravitational field.

\subsubsection{Semiclassical limit\label{hannay.sc}}
Following Hannay's work, Berry (1985) showed that
the Hannay angle emerges in the semiclassical limit of the geometric
phase.  The precise relation between the two is given by
    \begin{equation}                      \gamma_{n+1}(C) - \gamma_n(C)
   \xrightarrow[n \hbar  = I]{\hbar\rightarrow 0, n\rightarrow\infty}
                    \theta_g(I,C),\label{sc}\end{equation}
in which classical action is related to the quantum number by the
semiclassical quantization condition $I =  n \hbar$.  \eqref{sc}
conforms to the usual relation between quantum phases and
classical angles, a more familiar instance being the relation
$(E_{n+1} - E_{n})/\hbar \rightarrow \omega$ between energy levels
and frequencies, whose time integral connects the dynamical phase
and the dynamical angle. The result \eqref{sc} is derived  by
substituting the WBK approximation for the eigenstates
$\ket{n(\R)}$ into the formula \eqref{vecpot} for the quantum
vector potential $\A_n$.  One obtains thereby the expression
\eqref{Ac} for the classical vector potential. This analysis
extends to integrable systems in more than one dimension, for
which semiclassical eigenfunctions may be similarly determined.

On the other hand, for systems whose classical dynamics is
chaotic, the quantum geometric phase has no classical limit in general.
However, the vector field $\V_n$, when averaged over quantum
number, does have a classical limit (Robbins
and Berry 1992), which turns out to play a role in the classical
Born-Oppenheimer
approximation (cf Section~\ref{hannay.react}).  The classical limit
\begin{equation} 
V(E,\R) = -\frac{1}{2\partial_E \Omega(E,\R)} \partial_E \left(
   \partial_E\Omega(E,\R) \int_0^\infty \langle\gradR H_t(q,p,\R) \times \gradR
        H(q,p,\R)\rangle\,dt\right)\label{robber}\end{equation}
is expressed in terms of the time integral of an antisymmetric correlation
function of the parameter gradient of the Hamiltonian,
$\gradR H(q,p,\R)$.
$\Omega(E,\R) = \int_ {H(q,p,\R)\ge E} d^N q d^N p$
denotes the phase volume inside the energy shell $H = E$,
$\langle \cdot \rangle$ denotes the normalized
microcanonical average on the energy shell, and $\gradR H_t$ denotes the
function $\gradR H$ evolved for a time $t$ with the classical dynamics.
In
one dimension, \eqref{robber} is equivalent to \eqref{Vc}.

\subsubsection{Classical reaction forces\label{hannay.react}}
As in Section~\ref{bas.geomag}, we suppose the parameters $\R$ are no
longer externally prescribed, but instead are the coordinates of a slow
system to which the original one, the one-dimensional oscillator of
Section~\ref{hannay.cat}, is coupled.  Both systems are governed by
classical mechanics. We take their Hamiltonian to be  $P^2/2M +
H(q,p,\R)$, where $\P$ is the slow momentum. The Hannay angle describes
a holonomy in the fast dynamics induced by  the slow dynamics.  As in
the quantum case, there is a reciprocal effect on the slow dynamics
induced by the fast dynamics.  This effect is found by  averaging over
the fast motion. According to the adiabatic theorem, the action of the
fast system is nearly constant. In the lowest order approximation, we
treat the action as strictly constant.  Then the energy $H(I,\R)$  acts
as an potential in the slow dynamics, whose effective Hamiltonian is
given by $P^2/2M + H(I,\R)$.  This is the classical analogue of the
Born-Oppenheimer approximation.

First-order corrections account for the fluctuations of the  action
about its mean. These generate an effective vector potential in the slow
Hamiltonian, so that $P \rightarrow P - \A(I,\R)$.  The vector potential
is precisely the field \eqref{Ac} whose line integral gives the Hannay
angle.  This is classical geometric magnetism.

One example of geometric magnetism is the precession of a symmetric top
(Berry and Robbins 1993a). In this case, the top's axis is the slow
system, and the rotation  about this axis is the fast system.  Another
example from plasma physics is the drift of the guiding centre of electrons
in strong, slowly-varying magnetic fields (Littlejohn 1988). The
guiding centre, ie the centre of the instantaneous Larmor orbit,
constitutes the slow degrees of freedom (it remains fixed in constant
fields), and the motion  about the guiding centre, the fast degree of
freedom.

Geometric magnetism generalizes to higher dimensional fast systems,
both integrable and chaotic.  In the latter case, geometric magnetism
is partnered by an additional dissipative force, deterministic
friction; the slow dynamics is no longer conservative, as the chaotic
motion of the fast system acts as an effective heat bath.  The
magnetic and dissipative forces appear as the antisymmetric and
symmetric parts of a linear response tensor (Berry and Robbins 1993b,
Jarzynski 1993).

\subsection{Symmetry and kinematic geometric
phases\label{classical.kin}} Kinematic geometric phases arise in
systems with symmetry.  They describe the evolution in degrees of
freedom which the symmetry acts upon (eg, orientational degrees of
freedom in case of rotational symmetry).  Their structure depends only
on how the symmetry is realized, and not on the details of the
dynamics.  The geometric character of their evolution is exactly
enforced by the equations of motion, and does not depend on
assumptions of adiabaticity.  In the following sections, two different
treatments of kinematic geometric phases are described.

\subsubsection{Falling cats and swimming paramecia\label{kin.cats}}
Even with nothing to push against, bodies in free fall, such as cats and
platform divers, can change their orientation through internal
motions.  Perhaps counterintuitive at first, this phenomenon is a
direct conseqence of conservation of angular momentum, and can be
calculated as such.  The main difficulty in doing so is in the definition
of the body's orientation.  This problem does not arise for truly rigid
bodies, whose every configuration is related to a chosen reference by
a unique rotation.  But for an body which can bend and twist and
stretch, ie which has internal degrees of freedom, it is not clear
which configurations are ``upright'' and which ones are ``tilted''.

There are various prescriptions for determining orientation.  One is
to refer to the orientation of the principal axes of inertia.  Another
is through the Eckart frame, a common convention in molecular physics.
Both presciptions have particular advantages; neither is well defined
in all cases.

In fact, the structure of the solution emerges most clearly if the
definition of orientation is made more or less arbitrarily.  This is
the point of view taken by Shapere and Wilczek(1989b).  For
simplicity, suppose the body consists of a finite number $N$ of masses
$\ma$ at positions $\ra$.  No assumption is made concerning their
dynamics, except that they obey Newton's laws and that there are no
external forces.  By choosing a suitable reference frame, we may
assume that the centre of mass $\sum_\alpha \ma
\ra/\sum_\alpha \ma$ begins and remains at the origin.

By some means, certain configurations $(\R_1,\ldots, \R_N)$ are
designated to be ``upright''.  This designation must be consistent, in
that if two configurations are related by a rotation, then only one of
them can be upright.  It must also be complete, so that every
configuration can be made upright by applying a rotation.  If both
these conditions hold, then every configuration $(\r_1,\ldots,\r_n)$
can be obtained by applying a unique rotation $\calR$ to a unique
upright configuration $(\R_1,\ldots, \R_N)$.  The body's motion,
    \begin{equation}\ra(t) = {\calR}(t)\cdot\Ra(t),\quad \alpha = 1,\ldots,N
    \label{ra}\end{equation}
is then determined by the evolution of its orientation
${\calR}(t)$ and its internal dynamics $\R_1(t),\ldots, \R_N(t)$.
In the language of rigid body dynamics, $\R_\alpha(t)$ describes
the motion in the body frame, and $\r_\alpha(t)$ the motion in the
space frame.

An object fixed
in the space frame appears to rotate with angular velocity
$-\bOmega(t)$ in the body frame. $\bOmega(t)$ is defined by the
relation
        \begin{equation}\dot \calR\cdot\R =
        \calR\cdot(\bOmega\times\R),\label{omega1}\end{equation}
which can be made to hold for arbitrary vectors $\R$.

Regardless of the details of the dynamics, the total angular momentum $\l =
\sum_\alpha \ma \ra\times\va$ in the space frame must be conserved.
Conservation of angular momentum leads to the relation
    \begin{equation}\bOmega(t) = I^{-1}\cdot\left(
\L - \sum_\alpha \ma
\Ra\times\Va\right)\label{omega2}\end{equation} between the
angular velocity $\bOmega$, angular momentum $\L =
\calR^{-1}\cdot\l$,  velocities $\Va = \dot\Ra$, and inertia
tensor $I_{ij} = \sum_\alpha \ma(R_\alpha^2\delta_{ij} - R_{\alpha
i}R_{\alpha j})$ in the body frame.  Given the internal dynamics
$\Ra(t)$, the relations \eqref{omega1} and \eqref{omega2} may be
used to determine the orientation $\calR(t)$.

The case of zero angular momentum ($\l = \L = 0$) is particularly
interesting.  Then one can show that the orientation $\calR(t)$ does
not depend on the rate of the internal dynamics.  (One can, in effect,
multiply through by $dt$ in \eqref{omega1} and \eqref{omega2}.)  In
particular, if the internal motion is periodic along a cycle $C$, then
after one period it generates an overall rotation $\calR_g =
\calR(T)\calR(0)^{-1}$ of the body.  The rotation $\calR_g$ depends
only on the cycle $C$, and not on the rate it is traversed.  It is a
purely geometrical quantity, a nonabelian analogue (rotations do not
commute) of the Hannay angle.  It is the means by which a cat dropped
upside-down manages to land on its feet.

Like the other geometric
phases considered so far, this  can be described in terms of
holonomy on a fibre bundle.  The fibre bundle is the configuration
space of the masses $m_\alpha$, constrained to have centre of mass at the origin.
The base manifold consists of the distinct internal configurations, no
two of which can be made to coincide by a rotation. (Mathematically,
this is the quotient of the configuration space by the rotation
group.)  It can be parameterized by the upright configurations,
provided these are always well-defined.  The fibre consists of the
different possible orientations of a given internal configuration, and may be
parameterized by the rotation group.  A connection determines an
infinitesimal rotation $\delta \calR$ for an infinitesimal change of
internal configuration $\delta \Ra$, and the physically relevant
connection is determined by conservation of angular momentum.
Parallel transport round a closed
cycle $C$ in the base yields the holonomy $\calR_g$.  The curvature of
this connection, which describes the net rotation after an
infinitesimal cyclic change of shape, may be computed using a formula
analogous to \eqref{nonabV}.

The designation of upright configurations is arbitrary, and one can
always choose them differently.  It can be verified that the holonomies
$\calR_g$ do not depend intrinsically on this choice, and that the various quantities
which describe the orientation transform sensibly when a different
choice is made.  In particular, the curvature and connection transform
according to equations similar to \eqref{nonabgt2}.

A more complex example studied by Shapere and Wilczek(1989c) concerns swimming
through viscous fluids at low Reynolds number.  High viscosity and low
velocity means that bodies remain at rest unless acted upon by a force
-- a realization of Aristotelian mechanics!  Bodies locomote by
executing a series of internal motions, ie swimming strokes.
Conservation of linear and angular momentum determine the consequent
changes in position and orientation,  but now the motion of the fluid
must be taken into account. Solving the fluid dynamical problem leads to
a relation  similar to \eqref{omega2} for both the linear and angular velocity
in terms of rates of change of shape.  Applications in biology include
locomotion of microrganisms.

\subsubsection{Geometric phases and reduction\label{kin.red}}
Kinematic geometric phases can be framed in terms of the general
theory of symmetry in Hamiltonian systems, called {\it reduction}.
This formulation, extensively developed by J Marsden, R Montgomery and
coworkers (Marsden et al.\ 1991, Marsden 1992),
has led to a unified perspective on a
variety of geometric phase phenomena as well as a number of
interesting examples.

Dynamics is simplified by the presence of symmetry, and Hamiltonian
dyanamics especially so.  A symmetry always reduces the dimension of a
dynamical system by one, but in the Hamiltonian case, it reduces it by
two.  Indeed, when there is sufficient symmetry, the systems are
integrable, and the dynamics can be solved analytically (cf
Section~\ref{hannay.cat}).

The simplest symmetry is translation invariance in an ignorable
coordinate $q_N$.  Hamilton's equation $\dot p_N = -\partial
H/\partial q_N = 0$ (more generally, Noether's theorem) implies that
the conjugate momentum $p_N$ is conserved.  Fixing its value
$p_N=p_{N0}$ leads to a Hamiltonian
$H(q_1,\ldots,q_{N-1},p_1,\ldots,p_{N-1},p_{N0})$ in two fewer
variables.  Once their dynamics is obtained, the evolution of the
ignorable coordinate may be determined by integrating Hamilton's
equation $\dot q_N = \partial_{p_N} H(q_1,\ldots,q_{N-1},
p_1,\ldots,p_{n-1},p_{N0})$.  The theory of reduction extends this
analysis to general symmetries.  As in the simple example, the
analysis consists of two stages. The first is reduction, wherein the
conserved momenta are fixed, the ignorable coordinates ignored, and
Hamilton's equations expressed in terms of a reduced set of
coordinates.  In all but the simplest cases, finding these coordinates
is a nontrivial  task (Jacobi's ``elimination of the node'' in the
rotationally symmetric $N$-body problem is a representative example.)
The second state is reconstruction, wherein the evolution of the
ignorable coordinates is determined from the reduced dynamics.  It is
in the reconstruction stage than kinematic geometric phases can appear.  It turns
out that the equations of motion for the ignorable coordinates can be
separated into  dynamical and geometric terms.  The geometric term
depends only on the path described by the reduced dynamics, and not on
the rate it is traversed.  The dynamical term, by constrast, does
depend on the rate.  Moreover, the form of the geometrical term
depends only the realization of the symmetry, and not on the details
of the Hamiltonian.

In the example discussed in the preceding section, the flexible body
composed of $N$ point masses, the ignorable coordinates correspond to
$\calR$, the body's orientation,  the conserved momentum to the
space-fixed angular momentum $\l$, and the internal coordinates
$\Ra(t)$ to the reduced system.  The evolution of $\R(t)$ is obtained
by integrating the linear system \eqref{omega1}, in which angular
velocity $\bOmega(t)$ is given in terms of the reduced dynamics by
\eqref{omega2}.  Eq.~\eqref{omega2} is an example of a reconstruction
formula divided into its dynamical and geometric parts.  The first
term, $I^{-1}\cdot\L$, is the dynamic contribution.  The second,
$I^{-1}\cdot(\sum_\alpha \ma \Ra\times\Va)$, is the geometric
contribution.  As discussed above, its integral depends only on the
path $C$ described by the internal dynamics $\Ra(t)$.

R Montgomery (1991)  found an even simpler example in the dynamics of a free
rigid body.  Its configuration (with its centre of mass fixed at
the origin)
is described by a rotation $\calR$, for which it
will be convenient to use the Euler angle parameterization, $\calR =
\calR_{z}(\phi)\calR_{y}(\theta)\calR_{z}(\psi)$.
The space-fixed angular momentum $\l$ is
a constant of the motion, and without loss of generality we
can assume it lies in
the $z$ direction. The body-fixed angular momentum $\L(t) =
\calR(t)\cdot\l$ has polar coordinates $(\theta(t),\phi(t))$ on the sphere of
constant radius radius $L = |\L| = |\l|$.  Its dynamics is determined
by the Euler equations $\dot \L = \bOmega \times \L$, where $\bOmega =
I^{-1}\cdot\L$ is the angular velocity
and $I$ is the inertia tensor.  The Euler equations
describe the reduced dynamics, in which $\L$, or equivalently $\theta$ and
$\phi$, are the reduced coordinates.  The ignorable coordinate is
$\psi$.

Under the Euler equation, $\L$ executes a periodic motion
with period $T$.  The fact that $\L(T) = \L(0)$ implies
that rigid body returns after one period to its original orientation, up
to a rotation about the direction $\L(T)$.  Indeed, one has that
$\calR(T) = \calR(0)\calR_z(\Delta \psi) = \calR_{\L(T)}(\Delta \psi)
\calR(0)$, so that the amount of rotation,
$\Delta \psi = \psi(T) - \psi(0)$, is just the shift in the
ignorable coordinate $\psi$.  With a little algebra, the equation of
motion for $\psi$ may be written in the form
    \begin{equation}\dot \psi = 2E + \cos\theta \dot
    \phi,\label{rb}\end{equation}
where $E = \frac12 \L\cdot I^{-1}\cdot \L$ is the conserved
rotational energy.  \eqref{rb} is another example of a
reconstruction formula divided into dynamical and geometric
contributions.  In this case, the dynamical contribution
integrates to $2ET$, where as the geometric contribution gives
(modulo $2\pi$) the solid angle $\oint_C \cos \theta d\phi$
described by the body-fixed angular momentum $\L(t)$ over a single
period.

Another example, due to Alber and Marsden (1992), concerns phase shifts
in soliton collisions.  A spectacular property of certain nonlinear
partial differential equations, such as the Korteweg-de Vries equation
$U_t +6UU_x + U_{xxx} = 0$, is the existence of {\it solitons}.  These
are waveforms $U(x-vt)$ which propagate at a fixed velocity $v$ whilst
maintaining their shape, the nonlinearity in the wave dynamics just
balancing the dispersion.  When two solitons collide, they emerge
after a complicated but transient period of interaction essentially
intact.  The only signature of their collision is a shift $\Delta x$
in their position $x \pm \Delta x - vt$ ($+$ for the faster soliton,
$-$ for the slower one). This shift, which may be determined from an
asymptotic analysis of the exact solutions, can be interpreted as a
geometric phase. The same analysis applies to a variety of other
integrable nonlinear partial differetial equations, including the
nonlinear Schr\"odinger equation and the Klein-Gordan equation.

\subsubsection{Optimization and control\label{Opt}}
The study of kinematic geometric phases suggests a problem in
optimization. What is the most efficient internal motion for producing
a given holonomy?  As a specific example, consider a satellite
orbiting the earth in a fixed orientation, so that its angular
momentum about its centre of mass is zero. It is desired to change its
orientation using as little energy as possible.  The obvious method of
applying external torques (supplied by jets attached to satellite, for
example) has the drawback that any uncompensated force or torque
leaves the satellite in a state of translation or rotation.
The alternative is execute a
sequence of internal motions (eg, flywheels driven by motors, for
example); in the manner described above, these generate a change the
orientation without altering the total linear or angular momentum.

The problem of finding the most efficient cycle with a given holonomy
leads to a constrained variational problem in the internal degrees of
freedom.  It may also be formulated in terms of control theory. The
solution depends very much on how efficiency is measured.  In certain
circumstances, it may be advantageous to make some sizeable internal
motions initially in order to achieve a configuration for which the
geometric phase curvature form (cf Section~\ref{theory.hol.con}) is
large; then much smaller subsequent motions can produce relatively
large effects.  In other situations, it may be necessary to restrict
to small motions about some equilibrium configuration; in this case,
the variational problem may be analyzed in terms of infinitesimal
circuits.  The solutions often have a physical interpretation in terms
of motion in a magnetic, or more generally, a Yang-Mills field.
Shapere and Wilczek (1989d) find efficient swimming strokes through
viscous media in two- and three-dimensions under various conditions.
Marsden (1992) describes
a number of other examples.

\section{Optics\label{optics}}
Optics has provided fertile ground for studying the geometric phase.
(For a recent comprehensive survey of experimental activity, see
Bhandari (1996)).  Most investigations have concentrated on
polarization effects.  The polarization of a beam of monochromatic,
coherent light behaves like a quantum mechanical spin, and under cyclic
evolutions (and noncyclic ones as well) acquires a geometric phase
similar to that of a spinning particle in a magnetic field (cf
Section~\ref{bas.spin}) There are various ways to change polarization.
One is by varying the direction of propagation, as the polarization
vector must remain transverse to it (Tomita and Chaio 1986); in this
case the polarization has a spin-1, or vectorial character.  Another
is through optical activity induced by anisotropic media, in which
case the polarization has a spin-1/2 character (Pancharatnam 1956,
Bhandari and Samuel 1988, Bhandari 1988).

A beam of light passing through a cycle of polarization states may be
recombined with a second beam whose polarization has remained
unchanged. Their total phase difference (dynamical + geometrical) can
be determined, for example, from their interference pattern.  In
certain cases, the dynamical phase shift can be made to vanish.  In
others, it can be explicitly computed and subtracted from the total
phase.  In either case, one obtains a measurement of the geometric
phase.

\subsection{Geometric phase of coiled light\label{optics.coiled}}
In a straight, single-mode optical fibre of circular cross-section,
there is no preferred polarization; whatever the polarization
state of the source, it is preserved along the fibre.  This is no
longer so if the fibre is made to bend and wind; then the polarization
must change in order to remain transverse to the axis of propagation
$\that$.  Provided the radius of curvature $R = ||d\that/ds||^{-1}$ is
large compared to the wavelength $\lambda$ (here $s$ measures
distance along the fibre), then the polarization vector $\dhat(s)$ of the
electric displacement  $\D(s,\rho) = \Re e^{i(2\pi s/\lambda -
\omega t)} f(\rho)\dhat(s)$ (where $f(\rho)$ is the radial modal
profile) evolves along the fibre according to
    \begin{equation}\frac{d\dhat}{ ds} =
    -(\that.\dhat).\that.\label{dprime}\end{equation}
The corrections to this approximate solution of Maxwell's
equations  are of order $\lambda/R$.

For a general elliptical polarization, $\dhat$ is complex. Within
the adiabatic approximation, both
its real and imaginary parts $\dhat_r$ and $\dhat_i$ remain orthogonal
to the direction of propagation $\that$, and each satisfies
\eqref{dprime} separately.  As a function of $s$, $\that(s)$ describes a
curve $C$ on the unit sphere, and the vectors $\dhat_r(s)$ and
$\dhat_i(s)$ are tangent to the sphere at $\that(s)$.
Equation \eqref{dprime} is the condition for their parallel transport.
(Infinitesimal parallel transport of a tangent vector $\dhat$ at
$\that$ to a tangent vector $\dhat + d\dhat$ at $\that + d\that$ is
accomplished by rigidly transporting $\dhat$ to $\that + d\that$, and
then projecting the result onto the tangent plane there.)  If the
curve $C$ is made to close on itself, the vectors $\dhat_r$ and
$\dhat_i$ return rotated through an angle $\theta_g$ equal to the
solid angle subtended by $C$. $\theta_g$ is the geometric phase,
depending only on the geometry of $C$.

Tomita and Chiao (1986) observed this effect by introducing linearly
polarized light from a He-Ne laser into a length of helically wound
optical fibre whose initial and final directions coincided.  The
direction of linear polarization of the emergent beam was measured and
found to have turned through an angle equal to the solid angle
subtended by the cycle of directions.  The experiment was performed
with helices of both fixed and variable pitch, as well as for fibres
constrained to lie in the $xy$ plane.  In this last case, the cycle
of directions lies on the equator of the unit sphere, and the
subtended solid angle is a multiple of $2\pi$.  As predicted, the
direction of its linear polarization remains unchanged.

The original analysis of the Tomita-Chiao experiment was based on
quantum mechanical considerations, treating a photon propagating in
the fibre as a spin-one particle whose helicity $\s.\that$ is
constrained to be $\pm 1$.  (Classically this corresponds to left- and
right-circularly polarized light.)  Just as a spin with nonzero
magnetic moment tracks the direction of a slowly changing magnetic
field (cf Section~\ref{bas.spin}), the helicity tracks the direction
of propagation $\that$, and acquires a geometric phase equal to the
solid angle described by $\that$.  By decomposing a linear
polarization into a superposition of left- and right-circular
polarizations, the observed rotation of linear polaration may be
recovered.

A systematic derivation, however, requires more detailed analysis, and
this is most easily carried out within the context of classical
(rather than quantum) electromagnetic theory.  Indeed, the essential
features of such an analysis are contained in early work by SM~Rytov (1938)
and VV Vladimirsky (1941), who derived
and interpreted the parallel transport law \eqref{dprime} for
polarization in the geometric optics limit (see (Berry 1990) for a
discussion).  Strictly speaking, geometrical optics is not appropriate
for single-mode fibres (the transverse wavelength is comparable to the
fibre diameter).  Berry (1987a) has carried out a classical analysis
using the full Maxwell equations, and it is the conclusions of this
analysis which are presented above.

Kitano et al.\ (1987) performed an experiment related to Tomita and
Chiao's, in which the cylic change in the direction of propagation
was accomplished nonadiabatically through a series of discrete
reflections from mirrors.  These reflections reverse helicity instead
of conserving it, and the calculation of the geometric phase must be
appropriately modified.

\subsection{Pancharatnam's phase\label{optics.panch}}
When light passes through a homogeneous medium, its direction of
propagation remains constant. However, if the
medium is optically anisotropic, the polarization can change.
Such changes in polarization are
accompanied by a geometric phase, discovered in this
context by S.~Pancharatnam in
1956.  (See (Berry 1987b) for a contemporary discussion of
Pancharatnam's work.)
For definiteness we consider plane electromagnetic waves
propagating in the $z$ direction with fixed wavenumber $k$ and
frequency $\omega$.  The electric field $\E(\r,t) = \Re \sqrt{I}
e^{i(kz - \omega t)}\ket{A}$ is described by the intensity $I_A$ and
normalized polarization vector $\ket{A} = (A_x,A_y)$, whose complex
components determine the phase and relative intensity along $x$ and
$y$.  (Dirac notation is used for complex two-vectors to highlight an
analogy below with two-state systems.)

Given two such waves, what is their relative phase?  Unless their
polarization states $\ket{A}$ and $\ket{B}$ happen to be proportional,
it is not clear that the question is a sensible one.  Pancharatnam
found
a natural and physically relevant prescription for their phase difference
by considering the intensity $I_{A+B}$ of their superposition,
$\sqrt{I_A}\ket{A} + \sqrt{I_B}\ket{B}$.  This is given by
  \begin{equation}I_{A+B} = I_A + I_B
+ 2\sqrt{I_AI_B}\Re\braket {A} {B}.\label{intensity}\end{equation}
For fixed individual intensity $I_A$ and $I_B$, the superposed
intensity depends only on the interference term, and is maximized
when $\braket {A} {B} = 1$.   This is Pancharatnam's criterion for
polarizations being in phase. More generally, the relative phase
between two polarizations is taken to be $\arg \braket {A} {B}$
(this is well-defined unless $\ket {A}$ and $\ket {B}$ are
orthogonal).  It is the amount by which the phase of one
polarization must be retarded relative to the second in order that
the intensity of their superposition be a maximum.

Being in phase is not a transitive relation; if $\ket {A}$
is in phase with $\ket {B}$, and $\ket {B}$ in phase with $\ket {C}$,
then, in general, $\ket {C}$ is not in phase with $\ket {A}$.  The
nontrivial relative phase $\arg \braket {C} {A}$ has a simple
geometrical interpretation. To every polarization state $\ket {A}$ we
associate a vector $e_A$ on the unit sphere, namely its spin expectation
value $\me {A} {\bsigma} {A}$ (a unit vector) rotated through $\pi/2$
about the $y$-axis.  This is how polarization is represented, up to an
overall phase factor, on the {\it Poincar\'e sphere}.  The
$y$-rotation, a matter of convention, puts the left and right circular
polarizations at the north and south pole of the Poincar\'e sphere,
and the linear polarizations at the equator.  (In more mathematical
language, the Poincar\'e sphere is the projective Hilbert space for
two-state systems -- cf Section~\ref{theory.nonad}.)

One can show that the relative phase $\arg\braket {C} {A}$ is equal to
minus half the solid angle subtended by the spherical triangle $e_A e_B e_C$
on the Poincar\'e sphere, constructed by joining $A,B$ and $C$ with
arcs of great circles.  A direct demonstration (Pancharatnam's) uses
formulas from spherical trigonometry.  Another way is to realize that
a sequence of polarization states $\ket {\psi(t)}$
parallel-transported according to \eqref{HScon} along great circles on
the Poincar\'e sphere are, according to Pancharatnam's criterion, in
phase with each other.  Thus $\arg\braket {C} {A}$ is precisely the
geometric phase accumulated under parallel transport round $e_A e_B
e_C$, for which the half-solid-angle formula is discussed in
Section~\ref{bas.spin}.

To establish a connection between the geometry and an observable
effect, an analysis of the physical polarization dynamics is required.
Pancharatnam considered systems of {\it ideal polarizers}, or {\it
analyzers}, which completely transmit a polarization state $e_A$ and
completely absorb the state $-e_A$ orthogonal to it in such a way that
the incident and transmitted states are always in phase.  Ideal
polarizers are nearly approximated by very thin polarizing plates.
(Mathematically they are represented the hermitian projection operator
$\ket{A}\bra{A}$.)  Therefore, if a polarization cycle is generated by
ideal polarizers, one may conclude that the initial and final
polarizations differ by Pancharatnam's phase.

Polarization cycles can also be generated by {\it ideal retarders};
these are perfectly transparent devices which introduce a relative
phase difference between two given orthogonal polarization states
$e_A$ and $-e_A$.  A particular example is a quarter-wave plate, for
which the orthogonal states are linearly polarized and the phase
difference is $\pi/2$.  (Mathematically, they are represented by
$\exp(-i e_a\cdot\bsigma/2)$, a two-dimensional unitary matrix with
determinant one.)  Polarization cycles generated by ideal retarders produce
a dynamical phase in addition to Pancharatnam's phase.  The dynamical
phase is easily calculated, however.

Finally, one can consider optical activity generated by a transparent,
birefringent gyrotropic medium, which generates relative phase
differences between orthogonal polarization states whose orientation
varies with position.  When the variation is sufficiently slow over a
wavelength, a beam remains in a polarization state determined by
the locally uniform dielectric tensor.
Berry (1987b) applied an adiabatic analysis
to Maxwell's equations for this system and
calculated the phase for a
polarization cycle generated by a spatially periodic medium.  The
underlying structure is identical to a spin-1/2 particle in a slowly
changing magnetic field (Section~\ref{bas.spin}).  There is a
dynamical phase and a geometric phase, and the geometric phase is just
Pancharatnam's.

Observations of Pancharatnam's phase are based on two-beam
interferometry.  Such experiments were initiated by Pancharatnam
himself, though current versions are greatly improved with the
availability of laser sources.  Schematically, a plane wave with polarization
$\ket{A}$ is split into two coherent beams which propagate along the
arms of an interferometer.  Beam 1 propagates freely, while beam 2
passes through a sequence of optical elements which induce a
polarization cycle.  The beams are then recombined, and their total
relative phase determined from their interference.  Differences in
dynamical phases are determined and subtracted, and the geometric
phase subsequently measured.

Bhandari (1988) uses laser
interferometry methods to measure changes in the geometric phase,
rather than the geometric phase itself.  These are readily generated
(while keeping the relative dynamical phases fixed) by continuously
rotating a half-wave plate in the polarization cycle.

Kwiat and Chiao
(1991) observed Pancharatnam's phase in single-photon interference
experiments using a source of two-photon entangled states.  One
photon, the idler, passed through a narrow-band energy filter to a detector.
Observation of this photon fixed the energy of its partner, which was made
to pass through an interferometer as above.
Detection rates varied in accordance with the intensity, which
depends directly on
Pancharatnam's phase.  Various coincidence detectors were used to
confirm that genuine single-photon interference was being observed.

Finally, Berry and Klein (1996) perform a modern version of Pancharatnam's
experiment, measuring relative phase by direct observation of the
interference fringes.  They also investigate the relation between
discrete and continuous polarization cycles by varying the
number of optical components.

\section{Molecular and Condensed Matter Physics\label{applications}}

Several applications of the geometric phase to molecular spectroscopy
(Section~\ref{pseudo}), nuclear magnetic resonance
(Section~\ref{nmr}), and the quantum Hall effect (Section~\ref{qhe})
are discussed.  Amongst a number of applications which have been omitted
from discussion, we mention $\lambda$-doubling of rotational levels in
diatomic molecules (Moody et al.\ 1986), and atomic and reactive
molecular scattering (Mead and Truhlar 1979, Zygelman 1987, Mead 1992,
Kuppermann and Wu 1993).

\subsection{Pseudorotation in triatomic molecules\label{pseudo}}
Ordinarily, the angular degrees of freedom of a spinless system have
integer quantum numbers.  This is because, ordinarily, the angular
momentum $L_\phi$ conjugate to an angle $\phi$ is taken to be $-i\hbar
\partial_\phi$; its eigenfunctions $\exp (i\nu\phi)$ are single-valued
only if $\nu$ is an integer.  The Aharonov-Bohm effect (cf
Section~\ref{bas.geophase}) provides an exception.
In the presence of a magnetic flux line on the $z$ axis, with flux
$\alpha hc/e$, the azimuthal angular momentum of an electron is given
by $-i\hbar \partial_\phi + \alpha$, the constant term $\alpha$ coming from
the magnetic vector potential $\A^{(s)}(\rho,\phi,z) =
\alpha/\rho\,{\hat \bphi}$.  Therefore, the eigenvalues of $L_\phi$ are no longer
integral multiples of $\hbar$, but are given by instead by $(m +
\alpha)\hbar$, where $m$ is integral.

As discussed in Section~\ref{bas.geomag}, there emerges in the
Born-Oppenheimer approximation a molecular analogue of the
Aharonov-Bohm effect. For a given electric state $\ket{n}$,
the nuclear configuration space is threaded
by a magnetic flux concentrated at points where $\ket{n}$
is degenerate.  This flux appears in the nuclear
Hamiltonian as a geometric vector potential \eqref{vecpot}.
As
discussed by Mead (1992), a triatomic molecule $X_3$ composed of
identical atoms (eg, alkali metal trimers such as Na$_3$) provides a
simple class of examples.  A
striking consequence of the molecular Aharonov-Bohm effect
is the appearance of anomolous half-integer
quantum numbers. Their  existence has
now been confirmed experimentally (Dela\'cratez et al.\ 1986).  While this
phenomenon had been antipated by Longuet-Higgens et al.\ (1958), it is
most naturally understood in the context of the geometric phase.

The nuclear configuration space for a triatomic molecule has nine
degrees of freedom.  Six correspond to overall
translational and rotational motion, which
can be ignored in this discussion, leaving three internal vibrational degrees of
freedom.  These may be
parameterized by the internuclear distances $R_{12}$, $R_{23}$,
$R_{13}$.  Thus, the internal nuclear configuration space $Q$ is
the space of noncongruent triangles (the $R_{ij}$'s determining the
lengths of the triangle's sides) with vertices distinguished.

In the Born-Oppenheimer approximation,
the
electronic energy levels  and eigenstates
are regarded as functions on the internal configuration space $Q$.
Assuming the molecular Hamiltonian has time-reversal symmetry,
we expect degeneracies of the
electronic states to lie along one-dimensional curves in $Q$ (cf
Section~\ref{bas.tr}).

It turns out that some of these degeneracies
can be determined by symmetry considerations alone.  For equilateral
configurations $R_{12}= R_{23}= R_{13} = a$, the nuclear configuration
is invariant under the dihedral group $D_3$ of symmetries of the
equilateral triangle.
Therefore,
the electronic eigenstates $\ket{n(a,a,a)}$ must
transform according to an irreducible representation, or irrep, of
$D_3$ (Tinkham 1964).  The group $D_3$ has two one-dimensional irreps, associated with
states which are either invariant or else change sign under
the symmetry operations,
and a two-dimensional irrep, associated with
a degenerate pair of states which transform into each other under the
symmetry operations.
Therefore,
if $\ket{p(a,a,a)}$
transforms according to the two-dimensional
irrep, then $\ket{p(R_{12}, R_{23}, R_{13})}$
is necessarily degenerate along the curve $R_{12}=
R_{23}= R_{13} = a$ (with $a>0$) in $Q$.  Under parallel-transport
round this curve (cf \eqref{parallel}),  $\ket{p}$ changes sign.

One might expect the energy $E_p(R_{12}, R_{23}, R_{13})$ of this state
to assume its minimum in an equilateral configuration.  The
Jahn-Teller effect (Jahn and Teller 1937) shows this is not the case.
It turns out to be energetically favourable for the molecule to assume an asymmetrical
configuration, so that when it is in its $p$-electronic state,
the ground vibrational state of the molecule is distorted.

To analyze the vibrational motion, it is useful to
introduce cylindrical coordinates $(\rho,\phi,z)$ in $Q$.
The vertical coordinate $z$, given by $R_{12}^2 +  R_{23}^2 + R_{13}^2$,
determines the length scale; oscillations in $z$ describe
uniform dilations and contractions of the nuclei.  The radial
coordinate $\rho$, given by $2R_{12}^2 + (\sqrt{3}-1) R_{23}^2 - (\sqrt{3}+1) R_{13}^2$,
describes the degree of distortion; $\rho = 0$
corresponds to equilateral configurations.  Thus,
it is along the $z$ axis that the electronic state $\ket{p(\rho,\phi,z)}$
is degenerate.
The azimuthal angle $\phi$, given by $\tan^{-1} [{\sqrt{3}(R_{23}^2 -  R_{13}^2)/
          (2R_{12}^2 -  R_{23}^2 - R_{13}^2})]$,
describes the direction of the distortion.  Displacements in $\phi$
are called {\it pseudorotations}.  A $2\pi$-pseudorotation returns
the nuclei to their original positions, while a $2\pi/3$ rotation
induces a cyclic permutation amongst them.

A systematic calculation of the vibrational Hamiltonian $H_{\rm
vib}$ requires an explicit expression for the electronic energy
$E_p(R_{12}, R_{23}, R_{13})$ along with a careful derivation of
the kinetic energy in the $(\rho,\phi,z)$ coordinates.  Such a
treatment is described by Mead (1992); however, for this
discussion an approximate  form will suffice. The potential is
taken to be harmonic in $z$ and $\rho$ and independent of $\phi$,
and the kinetic energy to be of the standard form. Thus
\begin{equation} H_{\rm vib} = \frac{p_\rho^2}{ 2M} +
\frac{L_\phi^2}{2M\rho^2} +
   \frac{p_z^2}{2M} +  \frac12 M\omega_\rho^2 (\rho-\rho_0)^2
   + \frac12 M\omega_z^2 (z - z_0)^2,\label{Hvib}\end{equation}
where $(\rho_0,\phi_0)$ denotes the asymmetrical configuration
which minimizes $E_p$. The vibrational energy levels are
characterized by three quantum numbers $j,m,k$, for the respecive
degrees of freedom $\rho, \phi, z$.  They are given approximately
by
    \begin{equation}E(j,m,k) = (j+\frac12)\hbar\omega_\rho + \frac{(m\hbar)^2}{ 2I} +
                     (k+\frac12)\hbar\omega_z,\label{Ejmk}\end{equation}
where the azimuthal moment of inertia $I$ is taken to be
$M\rho_0^2$ (the approximation consists of ignoring the dependence
of $I$ on the radial quantum number $j$).

The presence of the electronic degeneracy along the $z$ axis
introduces a geometric vector potential $\A_p = \alpha {\bf \hat \bphi} /\rho$
into the nuclear Hamiltonian, which accounts for
the sign change of the electronic eigenstate under
parallel transport.  (In the chosen coordinates, $\alpha = \frac32$.)  Its
effect is to replace the pseudoangular momentum operator $L_\phi =
-i\hbar \partial_\phi$ by $-i\hbar \partial_\phi + \frac32 \hbar$,
just as in the magnetic Aharonov-Bohm effect.  Thus, the pseudorotational quantum number
$m$ in \eqref{Ejmk} is half-integral.

The half-integral quantization of pseudorotation was confirmed in a
systematic spectral analysis of the vibrational spectrum of Na$_3$ be
G Delacretez, ER Grant, RL Whetten, L W\"oste, and JW Zwanziger
(1986).  A sequence of rovibrational levels corresponding to radial
and pseudorotational motion were identified.  It was found that the
observed spectrum could be fitted to
\eqref{Ejmk} only
by assigning half-integer quantum numbers to the pseudoangular
momentum.  The result is an excellent agreement between observed and
calculated energy levels.


\subsection{Nuclear magnetic resonance\label{nmr}}

An rf magnetic field applied to polarized spins generates
transitions between their stationary states.  When the field is in
resonance with the spin precession frequency, the transition probabilities can
become large, even though the rf field may be weak.  Transitions
are manifested in oscillations of the spin magnetic moment, whose amplitude
and frequency can be measured with great accuracy.  This is the basis for nuclear spin
resonance experiments.  In nuclear magnetic resonance (NMR), the nuclear
spins are polarized by a static magnetic field.  In nuclear quadrupole
resonance (NQR), the spins are polarized by an electric field gradient
(in the ambient crytalline potential, for example) which couples to
the spin though the nuclear quadrupole moment.

Moody et al.\ (1986) and Cina (1986) pointed out that cyclic variations
in the rf  field would generate geometric phases in the nuclear
spins, which could be detected as shifts in the frequency
spectrum of the nuclear magnetic moment (this is an example of the third type
of experiment described in Section~\ref{bas.exper}).  Pursuing this
suggestion, several groups have carried out NMR studies of the geometric phase,
including its
adiabatic (Tycho 1987, Suter et al.\ 1987),  nonadiabatic
(Suter et al.\ 1988), and  nonabelian  (Zwanziger et al.\ 1990b) versions.
These experiments have not only confirmed theoretical predictions, but
have also brought to light interesting new phenomena in NMR
spectroscopy.

In an NMR experiment, the nuclear spins (let $s = 1/2$, say) are
initially polarized by a strong static magnetic field $B_z\zhat$; the spin
eigenstates $\ket{\pm}$ have energies $\pm\hbar\omega_0/2$, where
$\omega_0 = \gamma B_z$ is the Larmor frequency and $\gamma$ the
gyromagnetic ratio.  For nonstationary states, the magnetic moment
$\M(t) = \hbar\gamma\me {\psi(t)} {\bsigma} {\psi(t)}/2$
precesses about $\zhat$ at the Larmor frequency,
so that its frequency spectrum has a single peak at $\omega_0$.

A weak, circularly polarized rf field, $\B_{\rm rf}(t) = B_x \cos \omega t \xhat + B_y
\sin \omega t \yhat$, is then applied.
(Weak means that $B_{\rm rf}\ll B_z$).
The subsequent dynamics is most
easily analyzed in a rotating frame turning about
$\zhat$ with angular
frequency $\omega$.  There the effective magnetic field $\B_\rot$ is static,
and the Hamiltonian is given by
    \begin{equation} H_\rot = \frac12 \gamma \hbar \B_\rot\cdot \bsigma,\quad
 \B_\rot = \left(B_x \xhat + B_y \yhat + (B_z - \omega/\gamma)\zhat\right).
  \label{Hrot}\end{equation}
The spin eigenstates $\ket{\pm}_\rot$ in the rotating frame are
polarized along $\B_\rot$ with energies $\pm\hbar\omega_\rot/2$,
and the magnetic moment $\M_\rot(t)$ precesses about $\B_\rot$
with frequency $\omega_\rot$, where $\omega_\rot = \gamma
\B_\rot$. Near resonance, where $\omega\approx\omega_0$, the axis
of precession is nearly perpendicular to  $\zhat$ (even though
$B_{\rm rf}\ll B_z$).

The magnetic moment $\M(t)$ in the laboratory frame is obtained by turning
$\M_\rot(t)$ through an angle $-\omega t$ about $\zhat$.
The effect of the rf field is to superpose
a slow longitudinal nutation (of frequency $\omega_r$)
on the fast Larmor precession
of $\M(t)$ about $\zhat$ (of frequency $\omega_0$).
In consequence, the frequency spectrum of
the magnetic moment has additional peaks at
$\omega_0 \pm \omega_\rot$.

The Hamiltonian $H_\rot$ of \eqref{Hrot} depends parametrically on the
components of the rf and static magnetic fields, $B_x$, $B_y$ and  $B_z$,
and
vanishes for $B_x = B_y = 0$ and $B_z = \omega/\gamma$.
If these components are slowly modulated through a cycle $C$, with a
period $T$ much greater than the precession period $2\pi/\omega_\rot$ in
the rotating frame,
then the eigenstates $\ket{\pm}_\rot$ of $H_\rot$ evolve adiabatically,
and acquire with each modulation cycle
a geometric phase $\gamma_\pm(C)$
given by half the solid angle, $\alpha$, described by $C$ with
respect to $(0,0,\omega/\gamma)$ (cf
Section~\ref{bas.spin}).  These geometric phases produce
shifts $\mp \frac12 \hbar \alpha/T$ in the energies $\pm \frac12 \hbar
\omega_\rot$ of the unmodulated Hamiltonian $H_\rot$, and in turn shift the
precession frequency of the magnetic moment from
$\omega_\rot$ to $\omega_\rot - \alpha/T$.  (This shift is the
magnetic analogue of the daily precession of the
Foucault pendulum discussed in
Section~\ref{hannay.ex}.)

These frequency shifts were observed
by D Suter, GC
Chingas, RA Harris, and A Pines (1987) in the NMR response of proton spins
in a water/acetone sample.
The adiabatic cycle $C$ was produced by placing
the magnetization detector in a frame rotating with angular frequency
$\omega_\dett$ about $\zhat$.  In the detector frame, the effective
magnetic field turns about $\zhat$ with angular frequency $\delta =
\omega - \omega_\dett$; for small enough $\delta$, the spins adiabatically track
the turning field.
A variation of this experiment was subsequently performed
by D Suter, KT Mueller, and A Pines (1988), in which the geometric
phase was observed as a rotation in the spin-echoed
magnetic moment.  This technique enabled observation of the
individual geometric phases $\gamma_\pm$ rather
than the difference $\gamma_+ - \gamma_-$.
The spins were taken through nonadiabatic
cycles,  thereby incorporating the
Aharonov-Anandan phase (cf Section~\ref{theory.nonad}).

In a nuclear quadrupole resonance experiment, the spin polarization is induced by an
inhomogeneous electric field rather than a static magnetic field.
Gradients in the electric field couple to spin through
the nuclear electric quadrupole moment. The effective spin Hamiltonian
is of the form $H_Q = \hbar \omega_Q (\S\cdot\nhat)^2$, where $\S$ denotes
the dimensionless spin operators and $\nhat$ the symmetry axis of the
electric field gradient tensor (assumed to be axially symmetric).  The
eigenstates $\ket{m(\nhat)}$ are polarized along $\nhat$ with energies
$\hbar \omega_Q m^2$.  Note that these are doubly degenerate for $m\ne 0$.

Geometric phase effects in the NQR spectrum of $^{35}$Cl
(for which s = 3/2) were studied by R Tycko (1987).  In the experiment, a
coherent superposition of eigenstates of $H_Q$ was created at $t=0$ by an rf
pulse; the induced magnetic moment $\M(t)$ was then
observed and spectrally analyzed.
The spectrum of $H_Q$ consists of two degenerate doublets with
energies $\frac94 \hbar\omega_Q$  and $\frac14
\hbar\omega_Q$ (corresponding to $m = \pm 3/2$ and $m = \pm 1/2$
respectively), so that the magnetic moment spectrum has a
single peak at frequency $2\omega_Q$.
Adiabatic cycling of $H_Q$ was
achieved by rotating a crystalline sample of NaClO$_3$ with angular
frequency $\omega_R \ll \omega_Q$ about an axis
inclined to the symmetry direction $\nhat$.
With each rotation,
the spin
eigenstates acquire geometric phases $\beta_m$ (these may be computed explicitly),
which shift their energies by $\hbar
\beta_m \omega_R/2\pi$ and lift the double degeneracies.
For the particular geometry realized in Tycko's experiment, theory
predicts that the single peak in the magnetic moment
spectrum is split into three at  frequencies $2\omega_Q$ and $2\omega_Q \pm
\sqrt{3}\omega_R/\pi$.  Precisely this splitting was observed in the
experiment.

Since $H_Q$ has degenerate energy levels, one might expect
nonabelian geometric phases (cf
Section~\ref{theory.nonab}) to appear.  As shown by Tycko (1987) and
subsequently by A Zee (1988), the experiment can be analyzed
completely in terms of the ordinary (abelian) geometric phase.  The
nonabelian vector potentials $\Amb_{(\pm3/2)}$ and $\Amb_{(\pm1/2)}$
of \eqref{nonabA} (there is one  for each degenerate doublet of $H_Q$) can be
evaluated explicitly; they involve matrix elements of the spin angular
momentum between degenerate states.  For $m = \pm 3/2$, the
off-diagonal matrix elements vanish identically ($\S$ has nonzero
matrix elements only between states with $m$ differing by one), and
the diagonal matrix elements yield the ordinary geometric phases
$\beta_{\pm 3/2}$.  For $m=\pm 1/2$, these off-diagonal elements do
not vanish.  However, for the adiabatic cycles used in the experiment,
in which $\nhat$ is turned about a fixed axis, a basis
for the $m = \pm1/2$ doublet can found which diagonalizes $\Amb_{(\pm1/2)}$.
During the cycle there are no transitions between these basis vectors, and
each acquires an ordinary geometric phase $\beta_{\pm1/2}$.

Zee (1988)
shows that more general adiabatic cycles of $\nhat$  generate
genuine nonabelian geometric phases in the NQR spectrum.  JW
Zwanziger, M Koenig and A Pines (1990b) performed an experiment to
observe these nonabelian effects.  The more complicated adiabatic
cycles were generated by mounting the NaClO$_3$ crystal on a double
rotor.  Careful analysis predicts that the NQR resonance should be
split into five peaks instead of three.  This is precisely what was
observed, with excellent quantitative agreement between theory and
experiment.

\subsection{The quantum Hall effect\label{qhe}}
Charged particles in crossed uniform electric and magnetic fields
experience a net drift along the direction $\E\times\B$ normal to both
-- this is the basis for the {\it Hall effect}.  The Hall effect concerns
specifically the behaviour of electrons in a planar conductor to which
a perpendicular magnetic field $\B = B\zhat$ and a tangential electric
field $\E = E\xhat$ are applied.  The resulting current density has a
transverse component $j_y = \sigma_H E$.  $\sigma_H$ is the {\it Hall
conductance}.  Using classical theory, one obtains for large magnetic fields
that
     \begin{equation}\sigma^c_H \approx \frac{n e c}{
     B},\label{sigma_ch}\end{equation}
where $n$ is the two-dimensional electron density.

(Alternatively,
one can consider conductors of finite transverse extent, eg long thin
wires.  Then the current density is necessarily longitudinal, whereas
the electric field has a transverse component.  In this case it is
easier to compute the {\it Hall resistance} $\rho_H = E_y/j$.
Classical theory gives $\rho^c_H = B/nec$ for all magnetic fields, not
just large ones.  The Hall conductance and resistance are the
off-diagonal elements of the conductance and resistivity tensors
$\sigma$ and $\rho = \sigma^{-1}$, and the relation
between them, $\sigma_H =
\rho_H/(\rho_O^2 + \rho_H^2)$, depends on the
longitudinal or Ohmic resistivity $\rho_O$.)

For large magnetic fields and at low temperatures, the classical
prediction \eqref{sigma_ch} fails; what is actually observed is a
spectacular phenomenon, the quantum Hall effect.  As $B$ is increased,
instead of falling off monotonically as $1/B$, the Hall conductance
goes through a series of plateaus where its value is constant,
separated by intervals of monotonic decrease.  These constant values
are sample-independent, and satisfy with extremely high accuracy
(about one part in ten million) the quantization condition
    \begin{equation}\sigma_H = r \frac{e^2}{ h
    c}.\label{sigma_qh}\end{equation}
As $B$ increases, the coefficient $r$ takes on integer values, and
decreases with $B$ in unit steps until it reaches one.  This is
the regime of the integer quantum Hall effect (von Klitzing et
al.\ 1980).  As $B$ is increased still further, one enters the
regime of the fractional quantum Hall effect (Tsui et al.\ 1982);
$r$ takes on rational values ($<1$) with odd denominators.

The theory of the quantum Hall effect begins with the quantum mechanics
of two-dimensional electrons in magnetic fields.   The  one-electron
stationary states are the Landau levels $E_N = \hbar \omega_c(N+1/2), N
= 0,1,2,\ldots$, where $\omega_c = eB/mc$ is the classical cyclotron
frequency.  The Landau levels are highly degenerate; each can
accomodate $B/\Phi_0$ electrons per unit area ($\Phi_0 = hc/e$ is the
quantum of magnetic flux).   In real materials, the crystalline
potential, impurities, and electron-electron interactions all serve to
break this degeneracy.  However,
for sufficiently large magnetic fields, there is
negligible mixing between the Landau levels, so that each is split
by the perturbations into
$B/\Phi_0$ sublevels per unit area.

If the electron density in the sample is $n$, then the ground state has
the first $\nu = n \Phi_0/B$ Landau levels filled, the uppermost only
partially if $\nu$ has a fractional part. (Thus, the larger the
$B$-field, the fewer occupied Landau levels.)  The integer quantum Hall
effect can now be explained by making two assumptions: i) only a small
proportion of sublevels in each Landau level can carry current, and ii)
each filled Landau level contributes precisely $e^2/hc$ to the Hall
conductance.  Together, these imply that a nearly empty Landau level
contributes nothing to the Hall conductance, as the occupied sublevels
are unlikely to be the current-carrying ones, whereas, conversely, a
nearly full one contributes $e^2/hc$.  From these qualitative
considerations follows $\sigma_H =[\nu] e^2/hc$ ($[\nu]$ denotes
the integer nearest $\nu$), precisely \eqref{sigma_qh}, with $r =
[\nu]$.

The first assumption can be justified on the basis of localization
theory -- most of the sublevels are localized, whereas only extended
states can carry current.  For the second assumption, DJ Thouless,
M Kohmoto, MP Nightingale and M den Nijs (\tkn) found, in 1982, a
topological explanation.  In retrospect, their argument can be seen as
an application of the theory of the geometric phase. (Indeed, this
connection was drawn as early as 1983 by Simon.)

In outline, their analysis is as
follows.  We consider an electron in two dimensions in a uniform
magnetic field $B\zhat$ and periodic crystalline potential $V(x,y)$,
with periods $a$ and $b$ in $x$ and $y$.  Because the vector potential
$\A = (By/2,-Bx/2)$ is not periodic, neither is the Hamiltonian
$mv^2/2 + V(x,y)$, in which $\v = (\p +e\A/c)/m$ is the velocity.
However, the Hamiltonian {\it is} invariant under  the magnetic
translation operators $T_x$ and $T_y$, which shift $x$ and $y$ by $a$
and $b$ while leaving the velocity (rather than the momentum)
unchanged.  (Explicitly, $T_x = \exp(-iap_x/\hbar + i\pi Bay/\Phi_0)$
and $T_y = \exp(-ibp_y/\hbar - i\pi Bxb/\Phi_0)$.)

While the magnetic translation operators commute with the Hamiltonian,
they do not necessarily commute with each other.  In general, $T_x T_y =
\exp (2\pi i Bab/\Phi_0) T_y T_x$.  But if we take the flux $Bab$
through a unit cell to be a rational multiple $p/q$ of the magnetic flux
quantum $\Phi_0$, then $(T_x)^q$ -- a translation by $qa$ in $x$ -- and
$T_y$ do commute, and Bloch's theorem applies.  In this case, the energy
levels form a band structure, whose bands $\epsilon_n(\k)$ are periodic
over the first Brillouin zone $[0,2\pi/qa]\times[0,2\pi/b]$.  The
stationary states $\psi_{\k,n}(\r)$ are Bloch waves $\exp (i\k.\r)
u_{\k,n}(\r)$, where $u_{\k,n}(\r)$ is an eigenfunction of the
Hamiltonian
    \begin{equation}H(\k) = \frac12 m (-i\hbar\bnabla + e\A/c + \hbar\k)^2
    +V(x,y)\label{Hk}\end{equation}
with energy $\epsilon_{\k,n}$ satisfying the generalized periodic
boundary conditions $u_{\k,n}(x,y) =  \exp(i\pi
Bqay/\Phi_0)u_{\k,n}(x+qa,y) =  \exp(-i\pi
Bxb/\Phi_0)u_{\k,n}(x,y+b)$ over the unit cell.  (These are obeyed
rather than ordinary periodic boundary conditions because we are
dealing with magnetic rather than ordinary translation operators.)
For convenience we will sometimes denote the indices $n,\k$ by a
single label $\alpha$.

The Hall conductance can now be calculated perturbatively.  An applied
electric field $E\xhat$ generates perturbed eigenstates
$\ket{u'_\alpha} = \ket{u_\alpha} + \ket{\delta u_\alpha}$, and
induces transverse current densities $j_y = \sigma_H E$. To
lowest order in perturbation theory, these
are given by $-e \cdot 2\Re \me {u_\alpha} {v_y} {\delta
_\alpha}/A_0$.  Here $A_0$ is the area of the sample, and the
transverse velocity $v_y = -i\hbar\partial_y - eBx/c + \hbar k_y$
includes a contribution from the Bloch vector.  Note that $v_y$ may be
expressed as $\partial H/\partial \hbar k_y$.  From first-order
perturbation theory, $\braket {u_\beta} {\delta u_\alpha} = \me
{u_\beta} {-eEx} {u_\alpha}/(\epsilon(\alpha) - \epsilon(\beta)),
\beta\ne \alpha$.  Combining these results with the identity $[x,H] =
-i\hbar v_x$, we obtain the Kubo formula
  \begin{equation}\sigma_H = \frac{e^2}{ A_0 \hbar c} \sum_{\epsilon(\alpha) < \epsilon_F}
          \sum_{\beta\ne\alpha}
         \Im \frac{\me {u_\alpha} {\gradk H} {u_\beta}\times
         \me {u_\beta} {\gradk H} u_\alpha}{
               (\epsilon_\alpha - \epsilon_\beta)^2}
               \label{sh2}\end{equation}
for the Hall conductance of the ground state.  It includes
contributions from all states below the Fermi energy $\epsilon_F$.

Suppose we regard $H(\k)$ as a family of Hamiltonians parameterized by
the Bloch vector $\k$.  Under parallel transport round a closed cycle
$C$ in the first Brillouin zone, the eigenstate $\ket{u_{n,\k}}$
acquires a geometric phase $\gamma_{n,\k}(C)$, which may be expressed
(cf Section~\ref{bas.geophase}) as the integral of the flux field
$V_n(\k)  = \Im\me {\gradk u_{n,\k}} {\times}  {\gradk u_{n,\k}}$ over
the interior of $C$.  Amongst the various expressions for $V_n(\k)$
given in Section~\ref{bas.geophase}, we note that \eqref{alt}, which
involves matrix elements of the gradient of the Hamiltonian, is
contained precisely in the formula \eqref{sh2} for the Hall conductance.
This allows us to rewrite \eqref{sh2} as $\sigma_H = e^2/A_0\hbar c \sum_n
\sum_{\k} V_n(\k)$.  Finally, for samples large compared with the unit cell,
the sum $\sum_\k$ over Bloch vectors may
be replaced by an integral $A_0/4\pi^2 \int d^2\k$ over the first
Brillouin zone.  Thus we obtain
    \begin{equation}\sigma_H = \frac{e^2}{hc} \sum_n
    \sigma_n,\label{sh3}\end{equation}
where $\sigma_n = \int d^2\k V_n{\k}$ is the conductance (in units
of $e^2/hc$) of the $nth$ band.

\tkn show that $\sigma_n$ is an integer.  This follows by first
noting that $V_n(\k)$ is periodic over the Brillouin zone (even though
$\ket {u_{n,\k}}$ is not), so that the domain of the $\k$-space integral is
a closed torus $T_k$.  Indeed, one can show that $V_n(\k)$ is
given by $\Im \me
{\gradk \psi_{n,\k}} {\times} {\gradk \psi_{n,\k}}$ plus a perfect curl
(whose integral over the torus vanishes by Stokes' theorem).  The Bloch
states $\ket {\psi_{n,\k}}$ are periodic (up to a phase factor!) over the
first Brillouin zone, and constitute a complex line bundle (cf
Section~\ref{theory.hol.chern}) over $T_k$ with curvature $V_n(\k)$. The
integral over $T_k$ of the curvature over the torus is the first Chern
number, an integer.  For large magnetic fields, the bands coalesce into
nearly degenerate sublevels ($p$ per band) associated with the Landau
levels of a pure magnetic Hamiltonian.  One can show that sum of Chern
numbers from a given Landau level is equal to one.  This completes the
topological explanation for the quantization of Hall conductance.  \tkn
also calculate the Chern numbers for individual sublevels, which
are obtained from the  solutions of certain Diaphontine equations in $p$
and $q$.

\tkn was the first of a number of important applications of geometric
phase methods to the study of the quantum Hall effect. Several
subsequent works address the fractional effect, where electron-electron
interactions come into play.  Niu, Thouless and Wu (1984) show how the results of
\tkn could be generalized to accomodate them.  Arovas et al.\ (1984)
calculate properties of the elementary excitations
of Laughlin's  many-body fractional-quantum-Hall ground state (Laughlin 1983)
using geometric phase arguments. Concerning the
integer case, Arovas et al (1988) establish a topological criterion for
localization in terms of Chern numbers.  Wilkinson (1984) incorporates the
geometric phase into semiclassical calculations of the  single-particle
energy spectrum.

Finally, we mention studies by Avron and coworkers (Avron et al.\ 1988, Avron 1995)
of electron transport
in quantum systems.  While not specifically
concerned with the quantum Hall effect, they elucidate essentially similar
topological mechanisms.  One class of models consists of planar networks
of one-dimensional wires whose closed loops $C_i$ are threaded by
magnetic fluxes $\phi_i$ normal to the plane.  The physical properties
of the network are $\Phi_0$-periodic in the fluxes (fluxes differing by
an integral number of flux quanta are related by a gauge
transformation).  Fixing all but two of the fluxes, the stationary state
$\ket{\psi_n(\phi_i,\phi_j)}$ constitute a complex line bundle over the
two-torus of fluxes (the analogue of the first Brillouin zone). The
associated Chern number, the integral of the curvature, has the
following physical interpretation: it represents the average with
respect to $\phi_i$ of the number of electrons transported round $C_i$
as $\phi_j$ is increased adiabatically by one flux quantum.
\bigskip\bigskip

\noindent {\bf Glossary}
\bigskip

\begin{description}
\item{\it dynamical phase}\quad  The time integral of the energy of an evolving quantum
state divided by Planck's constant $\hbar$.
\smallskip
\item{\it geometric phase}\quad  A phase
factor accumulated by a cycled quantum eigenstate which depends only
on the geometry of its evolution.
\smallskip
\item{\it geometric magnetism}\quad Effective Lorentz-like force which appears
in the dynamics of a slow system coupled to a fast one after the fast
degrees of freedom have been averaged away.
\smallskip
\item{\it Hannay angle}\quad Displacement in the angle variables of
adiabatically cycled integrablesystems of purely geometric origin.
\smallskip
\item{\it holonomy}\quad Change in phase induced by parallel transport of a
quantum state around a closed curve.
\smallskip
\item{\it integrable system}\quad A Hamiltonian system with as many constants of
the motion in involution as there are degrees of freedom.
\smallskip
\item{\it parallel transport}\quad A prescription to determine the overall phase of an
evolving quantum state.
\smallskip
\item{\it reduction}\quad  General theory of Hamiltonian dynamics in the presence of symmetry.
\end{description}





\end{document}